\documentclass[a4paper,11pt]{article}
\pdfoutput=1 

\usepackage{jheppub} 

\usepackage[T1]{fontenc} 
\usepackage{subcaption}

\makeatletter
\DeclareRobustCommand*{\bfseries}{%
   \not@math@alphabet\bfseries\mathbf
   \fontseries\bfdefault\selectfont
   \boldmath
}
\makeatother

\newcommand{\tr}{\text{tr}}

\usepackage{tikz}
\usepackage{tikz-3dplot}
\usepackage{verbatim}

\usetikzlibrary{decorations.pathreplacing,decorations.markings}

\tikzset{
    >=stealth',
    punkt/.style={
           rectangle,
           rounded corners,
           draw=black, very thick,
           text width=6.5em,
           minimum height=2em,
           text centered},
    pil/.style={
           ->,
           thick,
           shorten <=2pt,
           shorten >=2pt,},
  on each segment/.style={
    decorate,
    decoration={
      show path construction,
      moveto code={},
      lineto code={
        \path [#1]
        (\tikzinputsegmentfirst) -- (\tikzinputsegmentlast);
      },
      curveto code={
        \path [#1] (\tikzinputsegmentfirst)
        .. controls
        (\tikzinputsegmentsupporta) and (\tikzinputsegmentsupportb)
        ..
        (\tikzinputsegmentlast);
      },
      closepath code={
        \path [#1]
        (\tikzinputsegmentfirst) -- (\tikzinputsegmentlast);
      },
    },
  },
  mid arrow/.style={postaction={decorate,decoration={
        markings,
        mark=at position .5 with {\arrow[#1]{stealth'}}
      }}}
}

\newcommand{\bra}[1]{\langle#1|} \newcommand{\ket}[1]{|#1\rangle}
\newcommand{\ketbra}[2]{|#1\rangle\!\langle#2|}
\newcommand{\braket}[2]{\langle#1|#2\rangle}  

\newtheorem{theorem}{Theorem}

\usetikzlibrary{decorations.pathmorphing}
\tikzset{snake it/.style={decorate, decoration=snake}}

\title{The holographic entropy zoo}

\author[a]{Alex May}
\author[a]{\!, Eliot Hijano}

\affiliation[a]{The University of British Columbia}

\emailAdd{may@phas.ubc.ca}
\emailAdd{ehijano@phas.ubc.ca}

\abstract{We study the holographic dual of a two parameter family of quantities known as the $\alpha$-$z$ divergences. Many familiar information theoretic quantities occur within this family, including the relative entropy, fidelity, and collision relative entropy. We find explicit bulk expressions for the boundary divergences to second order in a state perturbation whenever $\alpha$ is an integer and $z\geq0$, as well as when $z\in\{0,\infty\}$ and $\alpha\in \mathbb{R}$. Our results apply for perturbations around an arbitrary background state and in any dimension, under the assumption of the equality of bulk and boundary modular flows.}

\begin{document} 
\maketitle
\flushbottom

\section{Introduction}

\begin{figure}
\begin{center}
\begin{tikzpicture}[scale=0.65]

\node at (0,9.75) {$\alpha$-$z$ divergence};
\node at (0,9) {$D_{\alpha,z}(\rho||\sigma)$};

\node at (-7,5) {sandwiched relative entropy};
\node at (-7,4.25) {$D_{\alpha,\alpha}(\rho||\sigma)$};

\node at (7,5) {Petz relative entropy};
\node at (7,4.25) {$D_{\alpha,1}(\rho||\sigma)$};

\draw[blue,->] (-1,8.5)-- (-7,5.5) node [midway, above, sloped] (TextNode) {\small{$z=\alpha$}};
\draw[blue,->] (1,8.5) -- (7,5.5) node [midway, above, sloped] (TextNode) {\small{$z=1$}};

\draw[blue,->] (7,3.75) -- (1,-1.25) node [midway, above, sloped] (TextNode) {\small{$\alpha\rightarrow 1$}};
\draw[blue,->] (-7,3.75) -- (-1,-1.25) node [midway, above, sloped] (TextNode) {\small{$\alpha\rightarrow 1$}};

\draw[->,red] (5,4) -- (1,2.5) node [midway, above, sloped] (TextNode) {\small{$\sigma = \mathbb{I}$}};
\draw[->,red] (-5,4) -- (-1,2.5) node [midway, above, sloped] (TextNode) {\small{$\sigma = \mathbb{I}$}};

\node at (0,-1.75) {Umegaki relative entropy};
\node at (0,-2.5) {$S(\rho||\sigma)$};

\node at (0,2) {Renyi entropy};
\node at (0,1.25) {$S_\alpha(\rho)$};

\node at (9,-0.5) {von Neumann entropy};
\node at (9,-1.25) {$S(\rho)$};

\node at (-11,2) {log fidelity};
\node at (-11,1.25) {$-2\log F(\rho,\sigma)$};

\node at (-7,-1) {collision entropy};
\node at (-7,-1.75) {$D_{2,2}(\rho||\sigma)$};

\draw[blue,->] (-9,4) -- (-11,2.5)  node [midway, above, sloped] (TextNode) {\small{$\alpha=1/2$}};

\draw[blue,->] (-8,3.75) -- (-7,-0.5) node [midway, above, sloped] (TextNode) {\small{$\alpha=2$}};

\draw[->,blue] (0,0.75) to [out=-90,in=180] (6,-1);
\node at (4.5,-0.75) {\textcolor{blue}{\small{$\alpha \rightarrow 1$}}};

\end{tikzpicture}
\end{center}
\caption{Some members of the entropy zoo and their relations. The $\alpha$-$z$ divergences contain as special cases many common quantities appearing in quantum information theory. This figure is a simplified version of one produced by Faist \cite{entropyzoo}. Blue arrows indicate a choice of the parameters $\alpha,z$ has been made, red arrows that the reference state has been set to the identity matrix.}
\label{fig:entropyzoo}
\end{figure}
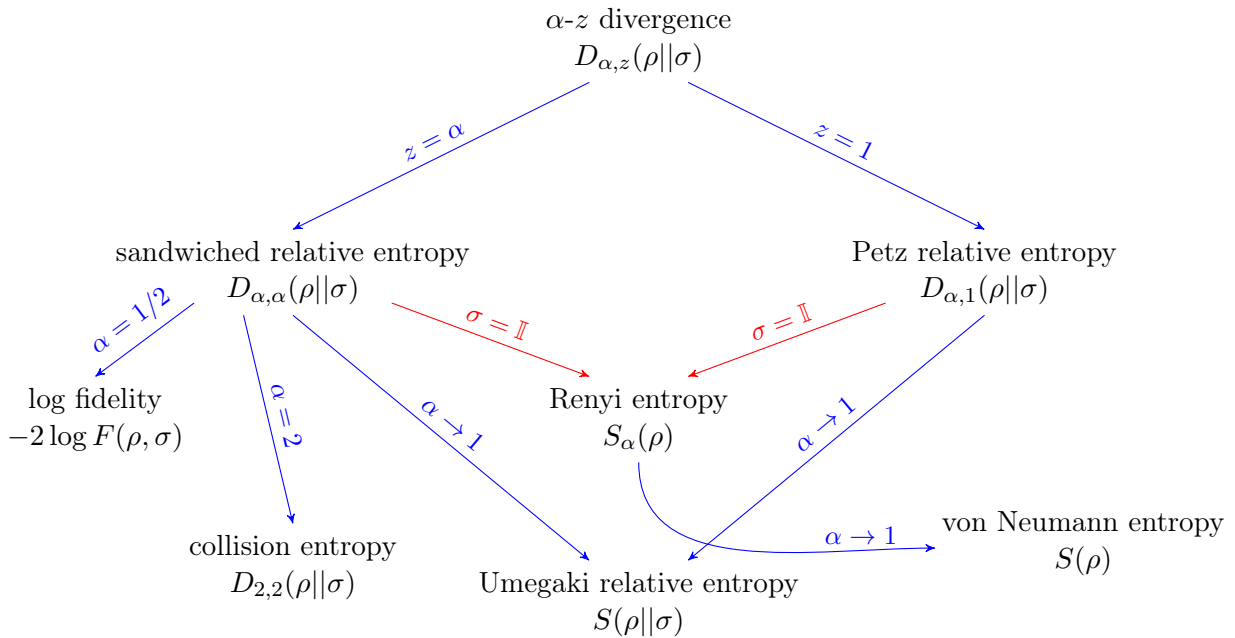

The understanding of the AdS/CFT correspondence \cite{Maldacena:1997re} has been deepened by the study of information theoretic quantities in conformal field theory. Most famously, the von Neumann entropy of a CFT subregion calculates the area of a minimal surface in AdS according to the Ryu-Takayanagi formula \cite{ryu2006holographic} or covariantly the HRRT formula \cite{hubeny2007covariant}. Other information theoretic quantities that have been studied in the context of AdS/CFT include the relative entropy \cite{lashkari2016canonical}, fidelity \cite{miyaji2015distance}, entanglement of purification \cite{takayanagi2017holographic,nguyen2018entanglement} and complexity \cite{brown2016holographic,susskind2016computational}.

The relative entropy and fidelity, as well as a number of other information theoretic quantities, occur as special cases of the $\alpha$-$z$ divergences \cite{audenaert2013alpha}. The $\alpha$-$z$ divergences are defined on two density matrices $\rho, \sigma$ according to
\begin{align}
D_{\alpha,z}(\rho||\sigma) = \frac{1}{\alpha-1}\log \tr \left( \left[ \sigma^{\frac{1-\alpha}{2z}}\rho^{\frac{\alpha}{z}}\sigma^{\frac{1-\alpha}{2z}} \right]^z\right).
\end{align}
The parameter ranges are restricted to $z \geq |\alpha-1|$ and $\alpha \geq 0$. We discuss the meaning of these quantities in section \ref{sec:divergences}, and note here that varying our choice of $\alpha$ and $z$ allows us to recover various standard quantities in information theory. Taking $\alpha=z$ recovers the sandwiched relative entropies \cite{muller2013quantum}, while setting $z=1$ gives the Petz relative entropies \cite{petz1986quasi}. We give an outline of the so called ``entropy zoo'' in figure \ref{fig:entropyzoo}, which portrays the relationships among the most commonly used information measures\footnote{We are following Phillipe Faist \cite{entropyzoo}, who produced an interesting poster after which our figure \ref{fig:entropyzoo} is modelled.}. 

It is currently unclear which information theoretic quantities should correspond to interesting gravitational observables. One goal of the present work is to systematically search for such quantities. Towards this, we study the $\alpha$-$z$ divergences leaving the choice of parameter values arbitrary. This allows us to address in one calculation a wide range of information theoretic quantities. Our calculation studies the $\alpha$-$z$ divergences perturbatively around a reference state. Considering $\rho$ a perturbation around $\sigma$, the first order term in $D_{\alpha,z}(\rho||\sigma)$ always vanishes. Consequently we study the second order term, which we label by $\chi_{\alpha,z}$ and refer to as the $\alpha$-$z$ susceptibilities. 

Throughout this work we make use of modular frequency space \cite{faulkner2017bulk}. In modular frequency space an operator $\mathcal{O}$ is expressed in terms of its modular frequency modes, which are defined using a density matrix $\sigma$ according to
\begin{align}
\hat{\mathcal{O}}_\omega \equiv \int_{-\infty}^{+\infty} ds\, e^{-is\omega} \sigma^{-is/2\pi} \hat{\mathcal{O}}\sigma^{is/2\pi}.
\end{align}
We find that the perturbative expansion of the $\alpha$-$z$ divergences simplifies when expressed in terms of the modular frequency modes of the state perturbation. This simplification, along with the equality of bulk and boundary modular flows \cite{jafferis2016relative} and the AdS/CFT extrapolate dictionary \cite{harlow2011operator,banks1998ads} allow us to find bulk duals of the susceptibilities in a simple way.

Our main result is an explicit bulk expression for the $\alpha$-$z$ susceptibilities whenever $\alpha=n$ is an integer. The bulk dual of the susceptibilities is expressed in terms of the modular frequency modes of the bulk fields and of their canonically conjugate momentum. Considering the case of a single bulk real scalar field $\Phi$ with conjugate momentum $\Pi$, we find that the susceptibility is
\begin{align}
\chi_{n,z} = \int_{\Sigma} d^dX \, \int_{-\infty}^{+\infty} d\omega \,iK_{n,z}(\omega) \left[\Phi(X)\Pi_{\omega}(X) - \Phi_{\omega}(X) \Pi(X) \right],
\end{align}
where $\Sigma$ is any spacelike slice of the entanglement wedge and the kernel $K_{n,z}(\omega)$ is a function we have determined explicitly. The field $\Phi$ refers to a perturbation to a certain background field, with the background fields fixed by the reference state $\sigma$. This expression is true for perturbations around an arbitrary background and in arbitrary dimensions. Taking the background to be the vacuum and setting $n=1$ we recover the bulk canonical energy, as established in earlier work \cite{lashkari2016canonical,faulkner2017nonlinear}.

The sandwiched relative entropies have been studied in the context of conformal field theory \cite{lashkari2014relative}. It was found that the sandwiched relative entropy for $\alpha=n$ an integer could be expressed as a $2n$ point function. Here we work perturbatively around a reference state and find that the second order term in the sandwiched relative entropy can be expressed as an integral over two point functions, for any $n$. Other related work includes \cite{bernamonti2018holographic}, which used holographic techniques to study the Petz relative entropies in conformal field theory. 

This paper is organized as follows. In section \ref{sec:divergences} we define the $\alpha$-$z$ divergences and state some inequalities they are known to satisfy. In section \ref{sec:holography} we introduce the holographic setting we work in. Section \ref{sec:holographicstates} defines the class of states we work with, while section \ref{sec:modular} reviews some needed details about modular flow and the relation between bulk and boundary flows. In section \ref{sec:holographicdivergences} we derive bulk expressions for the $\alpha$-$z$ susceptibilities. As a check on our work section \ref{sec:monocheck} uses this result to reproduce some known properties of the boundary susceptibilities starting from our bulk expressions. Appendices \ref{sec:eigexpansion} and \ref{sec:Fevenproof} deal with two technical aspects of our calculation; Appendix \ref{sec:equalityofdivergences} establishes the equality of bulk and boundary $\alpha$-$z$ divergences, extending the result of JLMS \cite{jafferis2016relative}.

\section{The \texorpdfstring{$\alpha$}{TEXT}-\texorpdfstring{$z$}{TEXT} divergences}\label{sec:divergences}

\subsection{Definition of divergence and some properties}

A divergence is defined as a function from two postive semidefinite operators to the real numbers, $D(\rho||\sigma)$, which satisfies the following axioms:
\begin{enumerate}
\item \textbf{Unitary invariance}: $D(\rho||\sigma) = D(U\rho U^\dagger||U\sigma U^\dagger)$.
\item \textbf{Normalization}: $D(\ketbra{0}{0}\,||\,\mathbb{I}/2)=1$.\footnote{This normalization is stated for the case of a single qubit, but this is enough to fix the normalization in general.}
\item \textbf{Order}: $D(\rho||\sigma)\geq 0$ when $\rho \geq \sigma$ and $D(\rho||\sigma)\leq 0$ when $\rho \leq \sigma$.\footnote{By $A\leq B$, we mean that $B-A$ is a positive semi-definite operator. Notice that it can never happen that two unit normalized density matrices have $\sigma > \rho$ or $\rho > \sigma$, as is clear from taking the trace. In fact for density matrices it is always the case that $D(\rho||\sigma)\geq 0$ with equality when $\sigma=\rho$.}
\item \textbf{Additivity}: $D(\rho_1\otimes \rho_2||\sigma_1\otimes \sigma_2) = D(\rho_1||\sigma_1)+D(\rho_2||\sigma_2)$.
\item \textbf{Generalized mean value}: There exists a continuous and strictly increasing function $g(x)$ defined for all $x \in \mathbb{R}$ such that
\begin{align}
[\tr(\rho_1)+\tr(\rho_2)]\,g(D(\rho_1\oplus \rho_2||\sigma_1\oplus\sigma_2)) = \tr(\rho_1)\,g(D(\rho_1||\sigma_1)) + \tr(\rho_2)\,g(D(\rho_2||\sigma_2)). \nonumber
\end{align}
\item \textbf{Continuity}: For $\rho\neq 0$ and $\text{supp}\, \rho \subseteq \text{supp}\, \sigma$, $D_{\alpha,z}(\rho||\sigma)$ is continuous in $\rho$, $\sigma\geq 0$.
\end{enumerate}
In the classical setting the relevant positive semidefinite operators are probability distribution interpreted as diagonal matrices, while in quantum mechanics this role is played by density matrices. In classical information theory the divergence of two probability distributions has an intuitive meaning. $D(P||Q)$ is the amount of information contained in an event $E$ about the variable $X$, where $P(X) = Q(X|E)$ \cite{renyi1961measures}. The above axioms formalize this intuition by specifying properties such a measure of information should have. In the quantum setting, a divergence is, at least heuristically, a measure of how distinguishable its two arguments are (this can be made precise in the context of quantum hypothesis testing \cite{mosonyi2015quantum}).

In the classical case the full set of all quantities satisfying these axioms is known \cite{renyi1961measures}. They are the Renyi divergences or Renyi relative entropies,
\begin{align}
D_\alpha(P||Q) = \frac{1}{\alpha-1} \log \sum_k  \left(\frac{p_k^\alpha}{q_k^{\alpha-1}}\right).
\end{align}
In the quantum case there are more divergences, owing to the non-commutativity of density matrices. In fact the full set of quantum divergences is not known.

All of the known quantum divergences are included in the $\alpha$-$z$ divergences \cite{audenaert2013alpha}. These are defined according to
\begin{align}\label{eq:alphazdef}
D_{\alpha,z}(\rho||\sigma) \equiv \frac{1}{\alpha-1} \log\tr\left( [\sigma^{\frac{1-\alpha}{2z}}\rho^{\frac{\alpha}{z}}\sigma^{\frac{1-\alpha}{2z}}]^z \right),
\end{align}
which are divergences in the sense of satisfying the axioms above whenever $z \geq |\alpha-1|$ and $\alpha\geq 0$. 

Two particular cases of the $\alpha$-$z$ divergences are well studied in the quantum information literature. The Petz relative entropies \cite{petz1986quasi} occur at $z=1$. They have been studied extensively in the context of quantum thermodynamics \cite{vinjanampathy2016quantum} and recently in the context of conformal field theory by using holographic techniques \cite{bernamonti2018holographic}. In both settings the Petz relative entropies can be thought of as free energies which must decrease under allowed operations, analogous to the second law of thermodynamics. The sandwiched Renyi relative entropies \cite{muller2013quantum} correspond to taking $z=\alpha$. These appear in the proof of the recoverability inequality, which has been employed in a holographic context to understand approximate bulk reconstruction \cite{cotler2017entanglement}. For simplicity, we will refer to these as the $\alpha$-$\alpha$ divergences and $\alpha$-$1$ divergences, respectively. Additionally, the combined quantity
\begin{align}
    \tilde{D}(\rho||\sigma) = 
    \begin{cases}
    D_{\alpha,1}(\rho||\sigma), & \alpha\in [0,1) \\
    D_{\alpha,\alpha}(\rho||\sigma), & (1,\infty)
    \end{cases}
\end{align}
has an operational meaning in terms of quantum hypothesis testing \cite{mosonyi2015quantum}.

The $\alpha$-$\alpha$ and $\alpha$-$1$ divergences reduce to the Umegaki relative entropy when $\alpha\rightarrow 1$, which reads
\begin{align}
S(\rho||\sigma) = \tr(\rho \log \rho - \rho\log \sigma). 
\end{align}
At second order the relative entropy of two CFT states is dual to the bulk canonical energy \cite{lashkari2016canonical}, which is the conserved quantity associated with evolution in modular time\footnote{The reader unfamiliar with modular time evolution may want to think of the canonical energy as associated with evolution in Rindler time, which coincides with modular time in the simplest case.}. At a technical level, the relative entropy being holographically dual to canonical energy has been useful in understanding how Einsteins equations in the bulk emerge from entanglement physics in the CFT \cite{faulkner2017nonlinear,Haehl:2017sot}.

Another familiar object occurs in the $\alpha$-$z$ divergences for $\alpha=z=1/2$, 
\begin{align}
D_{1/2,1/2}(\rho||\sigma) = -2\log F(\rho,\sigma),
\end{align}
where $F(\rho,\sigma)$ is the fidelity \cite{jozsa1994fidelity}. The fidelity is a natural generalization of the Hilbert space inner product on pure states $|\braket{\psi}{\phi}|^2$ to mixed states in that it is related to the inner product of purifications of $\rho$ and $\sigma$,
\begin{align}
    F(\rho,\sigma)=\underset{\ket{\phi},\ket{\psi}}{\max} |\braket{\psi}{\phi}|^2,
\end{align}
where the maximization is over purifications of $\rho$ and $\sigma$. In the context of holography the fidelity, expanded to second order in a state perturbation, has been argued to be dual to the volume of an extremal bulk surface \cite{miyaji2015distance}.

The $\alpha=z=2$ case is known as the collision relative entropy, 
\begin{align}
D_{2,2}(\rho||\sigma) = \log \text{tr}(\sigma^{-1/2}\rho\,\sigma^{-1/2}\rho).
\end{align}
This object evaluated on the particular states $\rho = \rho_{AB}$ and $\sigma = \mathcal{I}_A\otimes \rho_B$ is also sometimes called conditional collision entropy, $H_2(A|B) = D_{2,2}(\rho_{AB}|\mathcal{I}\otimes \rho_B)$, and appears commonly in quantum cryptography \cite{berta2013equality,dupuis2014one,renner2008security,dupuis2015entanglement,beigi2014quantum}.

Finally, another interesting case occurs when $\alpha=z\rightarrow \infty$, where
\begin{align}
    D_{max}(\rho||\sigma) = \inf \{\gamma : \rho \leq 2^\gamma \sigma \}.
\end{align}
This max relative entropy appears in the context of information theory in the single shot setting \cite{konig2009operational,datta2009min}

As a check on our later results we will prove a known property of the $\alpha$-$z$ susceptibilities starting from their bulk expressions. This property is given as the following theorems:
\begin{theorem}\label{thm:monotonealpharenyi}
The function $\alpha \rightarrow D_{\alpha,\alpha}(\rho||\sigma)$ is monotone increasing in $\alpha$.
\end{theorem}
\begin{theorem}\label{thm:monotonealphasandrenyi}
The function $\alpha \rightarrow D_{\alpha,1}(\rho||\sigma)$ is monotone increasing in $\alpha$.
\end{theorem}
Another interesting set of inequalities obeyed by divergences is monotonicity under the partial trace. The $\alpha$-$\alpha$ and $\alpha$-$1$ divergences are both known to be monotonic, as expressed in the following theorems.
\begin{theorem}\label{thm:tracemonotonealphalpha}
The $\alpha$-$\alpha$ divergence is decreasing under joint application of the partial trace to both of its arguments, $D_{\alpha,\alpha}(\rho_{A}||\sigma_{A})\leq D_{\alpha,\alpha}(\rho_{AB}||\sigma_{AB})$.
\end{theorem}
\begin{theorem}\label{thm:tracemonotonealphaone}
The $\alpha$-$1$ divergence is decreasing under joint application of the partial trace to both of its arguments, $D_{\alpha,1}(\rho_{A}||\sigma_{A})\leq D_{\alpha,1}(\rho_{AB}||\sigma_{AB})$.
\end{theorem}
Although the $\alpha$-$z$ divergences satisfy a number of interesting constraints \cite{audenaert2013alpha,carlen2018inequalities}, it is only the monotonicity in $\alpha$ and monotonicity under partial trace that we will explore in this work.

\subsection{The \texorpdfstring{$\alpha$}{TEXT}-\texorpdfstring{$z$}{TEXT} divergences in perturbation theory}\label{sec:suscep}

We study the $\alpha$-$z$ divergences $D_{\alpha,z}(\rho||\sigma)$ for $\rho$ close to a reference state $\sigma$. Consider a one parameter family of density matrices,
\begin{align}
\rho(\epsilon) = \sigma + \epsilon \, \delta^1\rho + \frac{\epsilon^2}{2} \, \delta^2 \rho +...
\end{align}
and expand the divergences according to
\begin{align}
D_{\alpha,z}(\rho(\epsilon)||\sigma) = D_{\alpha,z}(\sigma||\sigma) + \epsilon \left.\left(\frac{d}{d\epsilon} D_{\alpha,z}(\rho||\sigma)\right)\right|_{\epsilon=0} + \frac{\epsilon^2}{2}\left.\left(\frac{d^2}{d\epsilon^2}D_{\alpha,z}(\rho||\sigma)\right)\right|_{\epsilon=0} + O(\epsilon^3).
\end{align}
The $\alpha$-$z$ divergences are zero if and only if $\rho=\sigma$ and are always positive, which implies that the zeroth and first order terms vanish. This leaves the second order term as the leading one. We define
\begin{align}
\chi_{\alpha,z} \equiv \frac{1}{2}\left.\left(\frac{d^2}{d\epsilon^2}D_{\alpha,z}(\rho||\sigma)\right)\right|_{\epsilon=0},
\end{align}
and call this the \emph{$\alpha$-$z$ susceptibility}.

To study the susceptibility we recall the definition of the $\alpha$-$z$ divergence given as equation \ref{eq:alphazdef}. In Appendix \ref{sec:eigexpansion}, we discuss how to take derivatives of the combination of powers of density matrices that appears there by using some matrix identities. The result is that
\begin{align}\label{eq:knownform}
\chi_{\alpha,z}=\frac{z}{1-\alpha}\int d\sigma_a d\sigma_b  \frac{(\sigma_a^{\alpha/z}-\sigma_b^{\alpha/z})(\sigma_a^{\frac{1-\alpha}{z}}-\sigma_b^{\frac{1-\alpha}{z}})}{(\sigma_a-\sigma_b)(\sigma_a^{1/z}-\sigma_b^{1/z})} |\delta^1\rho_{ab}|^2.
\end{align}
The matrix elements of the perturbation $\delta^1\rho$ are in the eigenbasis of $\sigma$; that is $\delta^1\rho_{ab} = \bra{\sigma_a} \delta^1\rho \ket{\sigma_b}$ with $\sigma \ket{\sigma_a}=\sigma_a \ket{\sigma_a}$. 

To gain some intuition for equation \ref{eq:knownform}, it is helpful to consider a few special cases. Consider for example $\alpha=z$ corresponding to the sandwiched relative entropies. Then equation \ref{eq:knownform} simplifies to
\begin{align}\label{eq:alphalphaform}
\chi_{\alpha,\alpha}=\frac{\alpha}{1-\alpha}\int d\sigma_a d\sigma_b  \frac{(\sigma_a^{\frac{1-\alpha}{\alpha}}-\sigma_b^{\frac{1-\alpha}{\alpha}})}{(\sigma_a^{1/\alpha}-\sigma_b^{1/\alpha})} |\delta^1\rho_{ab}|^2.
\end{align}
Setting $\alpha=1/2$ we find
\begin{align}
\chi_{1/2,1/2}=\frac{1}{2} \int d\sigma_a d\sigma_b  \frac{2}{\sigma_a+\sigma_b} |\delta^1\rho_{ab}|^2.
\end{align}
while at $\alpha=2$ we have
\begin{align}
\chi_{2,2}=2\int d\sigma_a d\sigma_b  \frac{1}{\sqrt{\sigma_a\sigma_b}} |\delta^1\rho_{ab}|^2.
\end{align}
Since the arithmetic mean appears in the $\alpha=1/2$ case, and the geometric mean in the $\alpha=2$ case, we might expect $\chi_{\alpha,\alpha}$ to have the general form
\begin{align}\label{eq:averaging}
    \chi_{\alpha,\alpha} = \alpha \int d\sigma_a d\sigma_b \frac{1}{A_{\alpha}(\sigma_a,\sigma_b)} |\delta^1\rho_{ab}|^2,
\end{align}
where $A_\alpha(\sigma_a,\sigma_b)$ computes an average of the two eigenvalues. Indeed it is straightforward to check analytically from equation \ref{eq:alphalphaform} that this is the case\footnote{By $A_\alpha(\sigma_a,\sigma_b)$ computing an average we mean that it lies between its two arguments and satisfies $A_\alpha(\sigma_a,\sigma_a)=\sigma_a$.}. This is intuitive: for diagonal entries $\delta^1\rho_{aa}$ we would expect the contribution to the distinguishability of $\rho$ from $\sigma$ to be inversely proportional to $\sigma_a$ and proportional to $\delta^1\rho_{aa}$ (a small change to a big eigenvalue is hard to notice, while a big change to a small eigenvalue is easy to notice). Equation \ref{eq:averaging} shows that for off diagonal elements we weight the perturbation $\delta^1\rho_{ab}$ by an average of the two eigenvalues, and that the averaging function we use is fixed by the parameter $\alpha$.

Returning to \ref{eq:knownform}, we'd like to express the susceptibility in terms of the operators $\sigma$ and $\delta^1\rho$. Define 
\begin{align}
x \equiv \frac{1}{2\pi}\ln \frac{\sigma_b}{\sigma_a}
\end{align}
and use this to rewrite $\chi_{\alpha,z}$ as
\begin{align}\label{eq:chifromx}
\chi_{\alpha,z} = \frac{z}{1-\alpha} \int d\sigma_a d\sigma_b \frac{(1-e^{2\pi x\alpha/z})(1-e^{2\pi x \frac{1-\alpha}{z}})}{(1-e^{2\pi x})(1-e^{2\pi x/z})} \frac{1}{\sigma_a} \delta^1\rho_{ab} \delta^1\rho_{ba},
\end{align}
where we should remember that $x$ depends on $\sigma_a,\sigma_b$. Next, we split $\delta^1\rho$ in a sum over modes in a way that will let us recognize \ref{eq:chifromx} as a trace. In particular choose
\begin{align}\label{eq:hintedmod}
\delta^1\rho_{-\omega} \equiv \int_{-\infty}^{+\infty} ds \,e^{is\omega} \sigma^{-is/2\pi} \,\delta^1\rho \,\sigma^{is/2\pi}.
\end{align}
These \emph{modular frequency modes} of the operator $\delta^1\rho$ satisfy the completeness relation,
\begin{align}
\int_{-\infty}^{+\infty} d\omega \, \delta^1 \rho_{-\omega} = \delta^1\rho.
\end{align}
Notice that, as a consequence of definition \ref{eq:hintedmod}, 
\begin{align}\label{eq:modfrequencyfirstform}
(\delta^1\rho_{-\omega})_{ba} &= \delta(\omega-x) \delta^1\rho_{ba}.
\end{align}
Now we replace $\delta^1\rho_{ba}$ in \ref{eq:chifromx} with the integral over modular frequency modes. We can then interchange the order of the integrals over eigenvalues and integral over modes, and use that $\delta^1\rho_{-\omega}$ contains the delta function $\delta(\omega-x)$ to replace $x$ everywhere with $\omega$. We are left with 
\begin{align}
\chi_{\alpha,z} = \int_{-\infty}^{+\infty} d\omega F_{\alpha,z}(\omega) \int d\sigma_a d\sigma_b \frac{1}{\sigma_a} \delta^1\rho_{ab} (\delta^1\rho_{-\omega})_{ba},
\end{align}
where the function $F_{\alpha,z}(\omega)$ can be read from \ref{eq:chifromx},
\begin{align}
F_{n,z}(\omega) = \frac{z}{1-\alpha} \frac{(1-e^{2\pi \omega\alpha/z})(1-e^{2\pi \omega \frac{1-\alpha}{z}})}{(1-e^{2\pi \omega})(1-e^{2\pi \omega/z})}.
\end{align}
The double integral over eigenvalues is now recognized as a trace,
\begin{align} \label{eq:trueform}
\chi_{\alpha,z} = \int_{-\infty}^{+\infty} d\omega F_{\alpha,z}(\omega) \,\text{tr}(\sigma^{-1}\delta^1\rho\,\delta^1\rho_{-\omega}),
\end{align}
so that the susceptibilities are expressed in terms of the operators $\sigma, \delta^1\rho$. Alternatively, from definition \ref{eq:hintedmod} of the modular frequency modes we can express this in the modular time domain,
\begin{align}\label{eq:truetimeform}
\chi_{\alpha,z} = \int_{-\infty}^{+\infty} ds \tilde{F}_{\alpha,z}(s) \tr (\sigma^{-1}\delta^1\rho\, \sigma^{-is/2\pi}\delta^1\rho \,\sigma^{is/2\pi}),
\end{align}
where $\tilde{F}_{\alpha,z}(s)$ is the Fourier transform of $F_{\alpha,z}(\omega)$.

To find a bulk expression for these susceptibilities we will write them in terms of a time-ordered two point function (this is explained in detail at the end of section \ref{sec:holographicstates}). For this we need an alternative expression for $\chi_{\alpha,z}$. In particular we need that 
\begin{align}\label{eq:claimedotherform}
\chi_{\alpha,z} = \int_{-\infty}^{+\infty} d\omega F_{\alpha,z}(\omega) \,\text{tr}(\sigma^{-1}\delta^1\rho_{-\omega} \delta\rho),
\end{align}
which in the modular time domain reads
\begin{align}\label{eq:claimedtimeform}
\chi_{\alpha,z} = \int_{-\infty}^{+\infty} ds \tilde{F}_{\alpha,z}(s) \tr(\sigma^{-1}\sigma^{-is/2\pi}\,\delta^1\rho\, \sigma^{is/2\pi}\,\delta^1\rho).
\end{align}
Although expression \ref{eq:truetimeform} is always true, for \ref{eq:truetimeform} to be equal to \ref{eq:claimedtimeform} we need $\tilde{F}_{\alpha,z}(s)$ to be even\footnote{Alternatively, we could start with \ref{eq:truetimeform} and shift the $s$ contour in the complex plane by $2\pi i$ to obtain \ref{eq:claimedtimeform}. This is possible whenever $\tilde{F}_{\alpha,z}(s)$ is analytic in strip $0<\text{Im}(s)<2\pi$. Thus instead of studying where $\tilde{F}_{\alpha,z}(s)$ is even we could have studied its analyticity.}. In Appendix \ref{sec:Fevenproof} we establish that this is true for $\alpha \in \mathbb{N}$ and $z\geq 0$, and for $\alpha \in \mathbb{R}$ and $z\in\{0,\infty\}$. However, we show in section \ref{sec:noninteger} that $\chi_{\alpha,0}=0$, while at $z=\infty$, we find $\chi_{\alpha,\infty}$ is proportional to $\chi_{1,1}$ so that this case is already included in the integer $\alpha$ case. For this reason we focus on the case of integer $\alpha$ in what follows. 

Summarizing, we have
\begin{align}\label{eq:traces}
\chi_{n,z} &= \int_{-\infty}^{+\infty} d\omega F_{n,z}(\omega) \,\text{tr}(\sigma^{-1}\delta^1\rho_{-\omega} \delta^1\rho) = \int_{-\infty}^{+\infty} d\omega F_{n,z}(\omega) \,\text{tr}(\sigma^{-1}\delta^1\rho \,\delta^1\rho_{-\omega})
\end{align}
for $n\in \mathbb{N}$, where
\begin{align}\label{eq:Falphaz}
F_{n,z}(\omega) = \frac{z}{1-n} \frac{(1-e^{2\pi \omega n/z})(1-e^{2\pi \omega \frac{1-n}{z}})}{(1-e^{2\pi \omega})(1-e^{2\pi \omega/z})}.
\end{align}
These expressions do not make use of holography or quantum field theory; \ref{eq:traces} and \ref{eq:Falphaz} are statements only about the $\alpha$-$z$ divergences.

It is also sometimes of interest to define a metric on the space of quantum states starting from a susceptibility. The susceptibility is a quadratic function of the state perturbation, which allows us to promote it to a bilinear form on state perturbations. We do this by defining 
\begin{align}\label{eq:nzinfometric}
    \chi_{n,z}(\delta^1\rho_1,\delta^1\rho_2) \equiv \frac{1}{2}\left[\chi_{n,z}(\delta^1\rho_1 + \delta^1\rho_2) - \chi_{n,z}(\delta^1\rho_1) - \chi_{n,z}(\delta^1\rho_2) \right].
\end{align}
When $\alpha=z=1$, the bilinear form defined by the above is known as the Fisher information metric\footnote{Some authors use the convention that $\chi_{1/2,1/2}$ defines the Fisher information metric, our naming convention follows earlier usage in the high energy physics literature \cite{lashkari2016canonical}.} \cite{petz2011introduction}. Similarly, metrics defined from the Petz relative entropies \cite{hasegawa1993alpha} have been considered, and generally any divergence may be used to construct such a metric.

\section{The holographic setting}\label{sec:holography}

\subsection{Perturbations around path integral states}\label{sec:holographicstates}

In the context of Lorentzian AdS$_{d+1}$/CFT$_{d}$, we focus on a particular class of CFT states that are conveniently described using the Euclidean path integral \cite{Botta-Cantcheff:2015sav}. For these states, we will find that the $n$-$z$ susceptibilities can be written in terms of two point functions of the operators that were inserted into the path integral. We are interested in these particular states because they are understood to be dual to coherent bulk states associated with smooth geometries and field configurations \cite{skenderis2008real,christodoulou2016holographic,marolf2017euclidean}. Previous work involving excited states of this class concerns the emergence of bulk dynamics out of CFT entanglement thermodynamics \cite{faulkner2017nonlinear,Haehl:2017sot}.

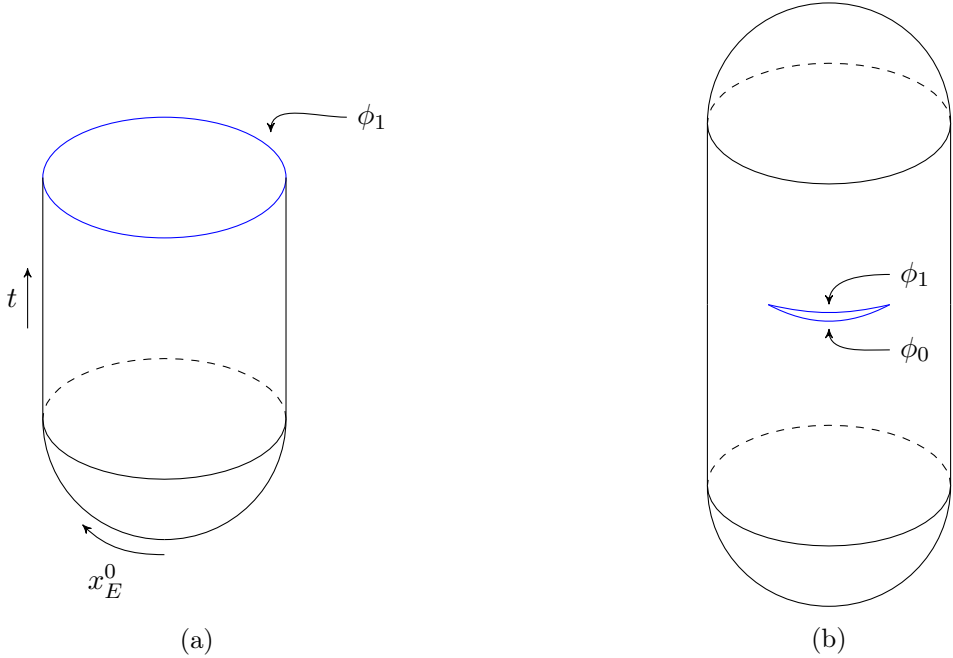
\begin{figure}
\centering
\begin{subfigure}{.45\textwidth}
\centering
\begin{tikzpicture}[rotate=90,scale=0.8]

\draw [dashed,domain=-90:90] plot ({cos(\x)}, {2*sin(\x)});
\draw [domain=90:270] plot ({cos(\x)}, {2*sin(\x)});

\draw[blue] (4,0) circle [x radius=1, y radius=2];

\draw (0,2) -- (4,2);
\draw (0,-2) -- (4,-2);

\draw (0,-2) to [out=180,in=-90] (-2,0);
\draw (0,2) to [out=180,in=90] (-2,0);

\node[left] at (2,2.25) {$t$};
\draw[->] (1.5,2.25) -- (2.5,2.25);

\draw[->] (-2.25,0) to [out=90,in=-135] (-1.75,1.35);
\node[below left] at (-2.25,0.5) {$x_E^0$};

\node at (6.75,0) {};

\node[right] at (5,-3) {$\phi_1$};
\draw[->] (5,-3) to [out=90,in=0] (4.75,-1.75);

\end{tikzpicture}
\caption{}
\label{fig:statepathintegral}
\end{subfigure}
\hfill
\begin{subfigure}{.45\textwidth}
\centering
\begin{tikzpicture}[rotate=90,scale=0.8]

\draw [dashed,domain=-90:90] plot ({cos(\x)}, {2*sin(\x)});
\draw [domain=90:270] plot ({cos(\x)}, {2*sin(\x)});

\draw (0,2) -- (3,2);
\draw (0,-2) -- (3,-2);

\draw (3,2) -- (6,2);
\draw (3,-2) -- (6,-2);

\draw (0,-2) to [out=180,in=-90] (-2,0);
\draw (0,2) to [out=180,in=90] (-2,0);

\draw [dashed,domain=-90:90] plot ({6+cos(\x)}, {2*sin(\x)});
\draw [domain=90:270] plot ({6+cos(\x)}, {2*sin(\x)});

\draw (6,-2) to [out=0,in=-90] (8,0);
\draw (6,2) to [out=0,in=90] (8,0);

\draw[blue] (3,1) to [out=-118,in=118] (3,-1);
\draw[blue] (3,1) to [out=-103,in=103] (3,-1);

\node[right] at (3.5,-1) {$\phi_1$};
\node[right] at (2.25,-1) {$\phi_0$};

\draw[->] (3.5,-1) to [out=90,in=0] (3,0); 
\draw[->] (2.25,-1) to [out=90,in=180] (2.6,0); 

\end{tikzpicture}
\caption{}
\label{fig:densitymatrixpathintegral}
\end{subfigure}
\caption{(a) The path integral given in equation \ref{eq:statepathintegral} that prepares the state $\ket{\Psi(t)}$. The cap on the lower end of the cylinder is the Euclidean path integral, while the cylinder is the Lorentzian part of the path integral. Fixing the boundary condition on the open end of the cylinder to a field configuration $\phi_1$ computes the amplitude $\braket{\phi_1}{\Psi}$. (b) The path integral given in equation \ref{eq:densitymatrixpathintegral} that prepares the density matrix on the subregion $R$. The path integral at left for preparing the state $\ket{\Psi(t)}$ has been doubled, and the two path integrals sewn together along $R^c$. Fixing boundary conditions above and below the cut along $R$ (shown in blue) computes the amplitude $\bra{\phi_1}\sigma_R\ket{\phi_0}$.}
\end{figure}

Recall that the vacuum state at time $x^0_E=0$ is described by the Euclidean path integral,
\begin{align}
\braket{\phi_1}{0} = \int^{\phi=\phi_1} \frac{[D\phi]}{N_\lambda}\exp\left(-\int_{-\infty}^{0} dx^E_0 \int d^{d-1}x \,\mathcal{L}_{CFT}[\phi]\right).
\end{align}
We can prepare another state in the CFT by turning on sources $\lambda_i$ for operators $\mathcal{O}_i$ in the CFT path integral,
\begin{align}\label{eq:pathintegralstates}
\braket{\phi_1}{\Psi(t=0)} &= \int^{\phi=\phi_1} \frac{[D\phi]}{N_\lambda}\exp\left(-\int_{-\infty}^{0} dx^E_0 \int d^{d-1}x \left(\mathcal{L}_{CFT}[\phi] + \sum_i\lambda_i \mathcal{O}_i\right) \right).
\end{align}
We will simplify our notation by writing
\begin{align}
S_E[\phi,\vec{\lambda}]=\int_{-\infty}^{0} dx^E_0 \int d^{d-1}x \left(\mathcal{L}_{CFT}[\phi] + \sum_i\lambda_i \mathcal{O}_i\right).
\end{align}
In order to prepare a state in the original CFT rather than one deformed by the operators $\mathcal{O}_i$ we require that $\lambda(x^0_E=0,x)=0$. We can attach a Lorentzian path integral at $x_E^0=0$ to implement time evolution,
\begin{align}\label{eq:statepathintegral}
\braket{\phi_1}{\Psi(t)} = \int D\phi \int_{\phi_L=\phi}^{\phi_L=\phi_1} D\phi_L \,e^{iS[\phi_L]} \int^{\phi_E=\phi} D\phi_E \,e^{-S[\phi_E,\vec{\lambda}]}.
\end{align}
The functional integration $D\phi_E$ is over the field on the Euclidean manifold, $D\phi_L$ is over the Lorentzian manifold, and $D\phi$ is the integration over the field configuration at their interface. The full path integral that prepares a state at some Lorentzian time is shown pictorially in figure \ref{fig:statepathintegral}. 

We will be interested in the state on a subregion $R$ of the CFT. The state on a subregion is expressed in terms of the reduced density matrix $\sigma_R$, which can also  be computed in terms of a path integral. Divide the fields $\phi$ into a part on $R$ and a part on the complement $R^c$ so that the full field configuration is specified by $(\phi,\phi^c)$. Then the reduced density matrix is
\begin{align}\label{eq:densitymatrixpathintegral}
\bra{\phi_1}\sigma_R(t)\ket{\phi_0} = &\int D\phi D\phi' D\phi''\int^{\phi'} D\phi_E' \,e^{-S[\phi_E',\vec{\lambda}]} \times \nonumber \\ 
&\int_{\phi'}^{(\phi_0,\phi)} D\phi_L'\,e^{iS[\phi_L']} \int_{(\phi_1,\phi)}^{\phi''} D\phi_L'' \,e^{iS[\phi_L'']} \int_{\phi''}D\phi_E'' \,e^{-S[\phi_E'',\vec{\lambda}]}.
\end{align}
We illustrate this path integral in figure \ref{fig:densitymatrixpathintegral}.

We will be interested in perturbing the density matrix $\sigma_R$. Working around a background defined by the sources $\vec{\lambda}$, we add some additional deformation $\epsilon \int \lambda \mathcal{O}$ to the Euclidean action so that the new density matrix is
\begin{align}
\bra{\phi_1}\rho_R(t)\ket{\phi_0} = &\int D\phi D\phi' D\phi''\int^{\phi'} D\phi_E' \,e^{-S[\phi_E',\vec{\lambda}]+\epsilon \int_{-\infty}^{0}dx_E^0 \int d^{d-1}x \lambda\mathcal{O}} \times \nonumber \\ 
&\int_{\phi'}^{(\phi_0,\phi)} D\phi_L'\,e^{iS[\phi_L']} \int_{(\phi_1,\phi)}^{\phi''} D\phi_L'' \,e^{iS[\phi_L'']} \int_{\phi''}D\phi_E'' \,e^{-S[\phi_E'',\vec{\lambda}]+\epsilon \int_{0}^{+\infty}dx_E^0 \int d^{d-1}x \lambda\mathcal{O}}.
\end{align}
Expanding the first and last exponentials to first order we find the perturbed density matrix is related to the old one by
\begin{align}
\rho &= \sigma + \epsilon \int_{-\infty}^{+\infty}dx_E^0\int d^{d-1}x \,\sigma \,\lambda(x_E^0,x) \mathcal{O}(x^0_E,x) + ...
\end{align}
This identifies
\begin{align}\label{eq:fieldssourced}
\delta^1\rho = \int d^{d} \mathbf{x}\, \sigma \,\lambda(\mathbf{x}) \mathcal{O}(\mathbf{x}) 
\end{align}
as the first order state perturbation, where we have defined $\mathbf{x}=(x_E^0,x)$. 

It is also useful to combine expression \ref{eq:fieldssourced} for the perturbation to a path integral state with the modular frequency transformation \ref{eq:hintedmod},
\begin{align}\label{eq:bulkmodesboundarymodes}
\delta^1\rho_{-\omega} = \int d^{d} \mathbf{x}\, \sigma \,\lambda(\mathbf{x}) \mathcal{O}_{-\omega}(\mathbf{x}).
\end{align}
This gives the modes of the state perturbation in terms of the modes of the operator insertions.

We're now ready to write the susceptibilities in terms of two point functions. Recall equation \ref{eq:traces},
\begin{align}
\chi_{n,z} &= \int_{-\infty}^{+\infty} d\omega F_{n,z}(\omega) \,\text{tr}(\sigma^{-1}\delta^1\rho_{-\omega} \delta^1\rho) = \int_{-\infty}^{+\infty} d\omega F_{n,z}(\omega) \,\text{tr}(\sigma^{-1}\delta^1\rho\, \delta^1\rho_{-\omega}).\nonumber 
\end{align}
This gives the $n$-$z$ susceptibilities in terms of the state perturbation $\delta^1\rho$ and its modes $\delta^1\rho_{-\omega}$. We now use \ref{eq:fieldssourced} and \ref{eq:bulkmodesboundarymodes} to write $\delta^1\rho$, $\delta^1\rho_{-\omega}$ in terms of the operator $\mathcal{O}$ and its modes $\mathcal{O}_{-\omega}$. However, we need to do this carefully to enforce time ordering. Use the first equality in \ref{eq:traces} when $x^0_E<y_E^0$,
\begin{align}
\chi_{n,z} = \int d\omega F_{n,z}(\omega) \int dx_E^0d^{d-1}x \,dy_E^0d^{d-1}y\lambda(x_E^0,x)\lambda(y_E^0,y) \,\tr (\sigma\mathcal{O}_{-\omega}(y_E^0,y)\mathcal{O}(x^0_E,x)).
\end{align}
While when $x^0_E>y_E^0$ use the second equality in \ref{eq:traces},
\begin{align}
\chi_{n,z} = \int d\omega F_{n,z}(\omega) \int dx_E^0d^{d-1}x \,dy_E^0d^{d-1}y\lambda(x_E^0,x)\lambda(y_E^0,y)\, \tr (\sigma\mathcal{O}(x^0_E,x)\mathcal{O}_{-\omega}(y_E^0,y)).
\end{align}
Taking these together we have
\begin{align}\label{eq:tordered2ptexpression}
\chi_{n,z}=\int_{-\infty}^{+\infty}d\omega F_{n,z}(\omega) \int d^d\mathbf{x}\,d^d\mathbf{y}\lambda(\mathbf{x})\lambda(\mathbf{y}) \,\langle \mathcal{O}(\mathbf{x})\mathcal{O}_{-\omega}(\mathbf{y}) \rangle.
\end{align}
This gives the $n$-$z$ susceptibilities in terms of an integral over the modular two point function. 

\subsection{Modular flow, free fields, and the modular extrapolate dictionary} \label{sec:modular}

From \ref{eq:tordered2ptexpression} we see that the $\alpha$-$z$ susceptibilities are conveniently expressed in terms of the modular frequency modes of the perturbing operators. Given this, it useful to recall a few facts about modular flow. The main goal of this section will be to obtain a bulk expression for the modular frequency mode $\mathcal{O}_{-\omega}$. Our summary here borrows from \cite{faulkner2017bulk}.

Given a region in space $R$ and a state on $R$ specified by a density matrix $\sigma_R$, there is a natural automorphism on the algebra of operators in $R$ given by
\begin{align}
\mathcal{O}(x) \rightarrow \mathcal{O}_s \equiv \sigma^{-is/2\pi}_R\mathcal{O}(x)\sigma^{is/2\pi}_R.
\end{align}
This operation is known as modular flow. The parameter $s$ is called modular time. In certain simple situations modular flow is geometric. For example in Rindler space, modular flow using the vacuum density matrix corresponds to Rindler time evolution.

It is useful to define the modular Hamiltonian according to
\begin{align}
H_R = -\log \sigma_R,
\end{align}
so that modular flow is Hamiltonian evolution with the modular Hamiltonian. Recently it has been argued that, in the AdS/CFT correspondence, the modular Hamiltonian for a CFT region $R$ is related in a simple way to the modular Hamiltonian of the corresponding entanglement wedge $W$ \cite{jafferis2016relative}. In particular,
\begin{align}\label{eq:modhamrelation}
H_R = \frac{\hat{A}}{4G} + H_{W},
\end{align}
which needs to be understood inside of expectation values, for instance we get that
\begin{align}
\tr(\sigma_R H_R) = \tr\left(\sigma_W\frac{\hat{A}}{4G}\right) + \tr\left(\sigma_W H_{W}\right).
\end{align}
The operator $\hat{A}$ measures the area of the RT surface. Importantly, $\hat{A}$ commutes with all the operators supported on any spacelike slice of the entanglement wedge. 
This has been used to establish the equality of bulk and boundary relative entropy for pairs of states that share an area operator. We point out an extension of this result to the $\alpha$-$z$ divergences in Appendix \ref{sec:equalityofdivergences}. This has also been used to establish the equality of bulk and boundary modular flows,
\begin{align}
\sigma_R^{-is/2\pi}\mathcal{O} \sigma_R^{is/2\pi} = \sigma_W^{-is/2\pi}\mathcal{O} \sigma_W^{is/2\pi},
\end{align}
which again needs to be understood inside of expectation values, and where on the left we consider $\mathcal{O}$ as a boundary operator and on the right as the same operator represented in the bulk.

The modular frequency modes, already introduced above, are defined by the Fourier transform,
\begin{align}
\mathcal{O}_\omega(x) \equiv \int_{-\infty}^{+\infty} ds\, e^{-is\omega} \mathcal{O}_s(x) = \int_{-\infty}^{+\infty} ds\, e^{-is\omega} \sigma^{-is/2\pi}\mathcal{O}(x)\sigma^{is/2\pi}
\end{align}
The correlator $\langle(\mathcal{O}_1)_s\mathcal{O}_2\rangle$ is analytic in the strip $0<\text{Im}(s)<2\pi$. From this fact we can derive the KMS condition,
\begin{align}
\langle (\mathcal{O}_1)_s \mathcal{O}_2\rangle = \langle \mathcal{O}_2 (\mathcal{O}_1)_{s+2\pi i}\rangle.
\end{align}
In the frequency domain the KMS condition allows us to prove that
\begin{align}\label{eq:freqKMS}
\langle (\mathcal{O}_1)_\omega \mathcal{O}_2 \rangle = n_\omega \langle [\mathcal{O}_2,(\mathcal{O}_1)_\omega] \rangle \,\,\,\,\, \text{with}\,\,\,\,\,n_\omega = \frac{1}{e^{2\pi\omega}-1},
\end{align}
which we will find useful in that it relates expectation values to commutators.

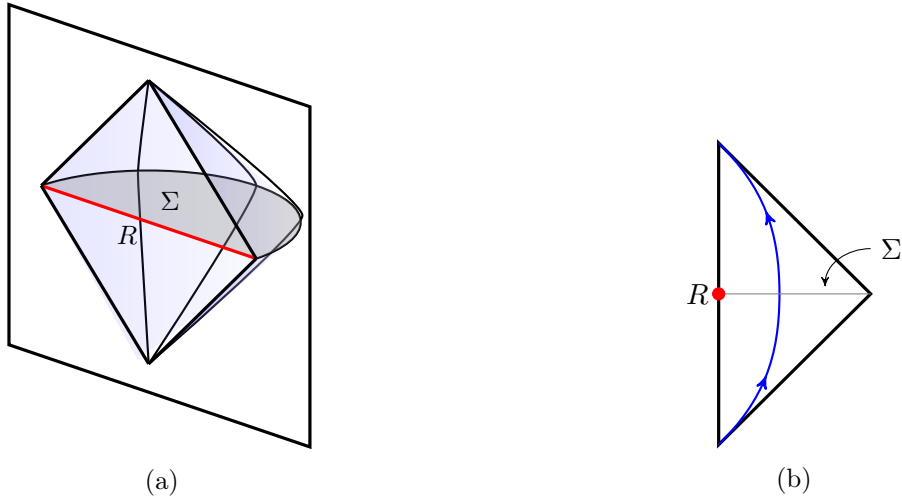
\begin{figure}
\centering
\begin{subfigure}{.45\textwidth}
\centering
\begin{tikzpicture}[scale=2,tdplot_main_coords]
\coordinate (O) at (1.3,0,1.2+1);


\draw [thick] plot [smooth, tension=0.2] coordinates { (1.3,0,1.2-1) (1.3,1,1.2) (1.3,0,1.2+1)};

\draw [thick] plot [smooth, tension=0.2] coordinates { (1.3,0,1.2-1) (1.3+0.43,1,1.2-0.1) (1.3,0,1.2+1)};

\draw [thick] plot [smooth, tension=0.2] coordinates { (1.3,0,1.2-1) (1.3-1.1,1,1.2-0.2) (1.3,0,1.2+1)};

\shade[left color=blue,right color=white,opacity=0.1] (1.3-1,0,1.2) arc[radius=1, start angle=180, end angle=25]  -- (1.2,0,1.2-1) -- cycle;

\shade[left color=blue,right color=white,opacity=0.3] (1.3+1,0,1.2) arc[radius=1, start angle=0, end angle=25]  -- (1.3,0,1.2-1) -- cycle;

\draw[thick] (1.3+1,0,1.2) arc[radius=1, start angle=0, end angle=180];
\fill[fill=gray,opacity=0.3] (1.3+1,0,1.2) arc[radius=1, start angle=0, end angle=180];
\draw[draw=red,very thick] (1.3+1,0,1.2) -- (1.3-1,0,1.2) node[pos=0.6, below] {\small $R$} node[pos=0.4, above=3] {\small{$\Sigma$}};

\shade[left color=blue,right color=white,opacity=0.1] (1.3-1,0,1.2) arc[radius=1, start angle=180, end angle=65]  -- (1.3,0,1.2+1) -- cycle;

\shade[left color=blue,right color=white,opacity=0.1] (1.3+1,0,1.2) arc[radius=1, start angle=0, end angle=65]  -- (1.3,0,1.2+1) -- cycle;

\draw[very thick, black] (1.3+1,0,1.2) -- (1.3,0,1.2+1);
\draw[very thick, black] (1.3+1,0,1.2) -- (1.3,0,1.2-1);
\draw[very thick, black] (1.3-1,0,1.2) -- (1.3,0,1.2+1);
\draw[very thick, black] (1.3-1,0,1.2) -- (1.3,0,1.2-1);

    \draw[
        draw=black,very thick
    ]          (0,0,0)
            -- (2.8,0,0)
            -- (2.8,0,2.4)
            -- (0,0,2.4)
            -- cycle;

\end{tikzpicture}
\caption{}
\end{subfigure}
\hfill
\begin{subfigure}{.45\textwidth}
\centering
\begin{tikzpicture}[scale=2]

\draw[very thick] (0,0) -- (1,1) -- (0,2) -- cycle;
\draw[thick,blue,postaction={on each segment={mid arrow}}] (0,0) to [out=45,in=-90] (0.4,1);
\draw[thick,blue,postaction={on each segment={mid arrow}}] (0.4,1) to [out=90,in=-45] (0,2);

\draw[gray] (0,1) -- (1,1);

\draw[red,fill=red] (0,1) circle (0.04);
\node[left] at (0,1) {$R$};

\draw[->] (1,1.3) to [out=180,in=90] (0.7,1.05);
\node[right] at (1,1.3) {$\Sigma$};

\node at (0,2.85) {};

\end{tikzpicture}
\caption{}
\end{subfigure}

\caption{(a) The boundary region $R$ (shown in red) and its associated entanglement wedge $W$ (shown as the blue shaded region). $\Sigma$ (shown in grey) is a spacelike slice of the wedge that is anchored to $R$ and the extremal surface. (b) A lower dimensional slice of the entanglement wedge. When the reference state is the vacuum, modular flow moves operators along the blue trajectory.}
\label{fig:entanglementwedge}
\end{figure}

We will work mainly in terms of a bulk scalar field, although we also show in section \ref{sec:gravityderivation} how to generalize our derivation to the spin 2 case. We will be interested in the configuration of this bulk scalar field on a spacelike slice $\Sigma$ of the entanglement wedge. Our set-up is shown in figure \ref{fig:entanglementwedge}.   For our bulk scalar, we work around some background field configuration that we label $\Phi_{\text{ref}}(X)$. The bulk field perturbation we call $\Phi(X)$. Working at leading order in $1/N$, we treat this perturbation as a free field. The scalar field operator $\hat{\Phi}$ is related to $\Phi_{\text{ref}}(X)$ and $\Phi(X)$ according to
\begin{align}
\Phi_{\text{ref}}(X) &= \tr(\sigma \,\hat{\Phi}(X)) \label{eq:phiphihatrelation} \\
\Phi(X) &= \tr(\delta^1\rho \,\hat{\Phi}(X)). \label{eq:phiphihatrelationp2}
\end{align}
Similarly we introduce the conjugate momentum operator $\hat{\Pi}(X)$, which satisfies the usual commutation relations with $\hat{\Phi}(X)$. Note that in equations \ref{eq:phiphihatrelation} and \ref{eq:phiphihatrelationp2} we are considering the operator $\hat{\Phi}(X)$ as living in the boundary Hilbert space, which we may do whenever $X$ is in the entanglement wedge of $R$ \cite{cotler2017entanglement,dong2016reconstruction,almheiri2015bulk}.

\tdplotsetmaincoords{70}{45}

To relate $\hat{\Phi}(X)$ to the boundary operator insertions, recall the extrapolate dictionary,
\begin{align}
\lim_{z\rightarrow 0} z^{-\Delta_{+}}\hat{\Phi}(x,z) = \mathcal{O}(x).
\end{align}
Notice that equality of bulk and boundary modular flows implies,
\begin{align}
\lim_{z\rightarrow 0} z^{-\Delta_{+}}\sigma_W^{-is}\hat{\Phi}(x,z)\sigma_W^{is} = \sigma_R^{-is/2\pi}\mathcal{O}(x)\sigma_R^{is/2\pi}.
\end{align}
We can integrate both sides of this expression against $e^{-is\omega}$ to arrive at the \emph{modular extrapolate dictionary},
\begin{align}\label{eq:modularextrapolate1}
\lim_{z\rightarrow 0} z^{-\Delta_{+}}\Phi_\omega(x,z) = \mathcal{O}_\omega(x).
\end{align}
Because our operator insertions are in the Euclidean path integral, while the extrapolate dictionary relates bulk fields and operators on the Lorentzian boundary, it is necessary to further modular evolve expression \ref{eq:modularextrapolate1} above into imaginary time. We define the operator
\begin{align}
    \hat{\Phi}_{\omega,i\tau}(X) \equiv \sigma^{\tau/2\pi} \hat{\Phi}_\omega(X)\sigma^{-\tau/2\pi}. 
\end{align}
Then our final version of the extrapolate dictionary is
\begin{align}\label{eq:modularextrapolate2}
\lim_{z\rightarrow 0} z^{-\Delta_{+}}\Phi_{\omega,i\tau}(x,z) = \mathcal{O}_{\omega}(x,\tau).
\end{align}
It will also be useful to re-express the modular frequency modes as linear combinations of the field operators. In particular, since $\hat{\Phi}(X)$ and $\hat{\Pi}(X)$ form a basis of operators for the algebra on $\Sigma$, there must be coefficients $C(X,Y,\tau)$ and $D(X,Y,\tau)$ such that 
\begin{align}
\hat{\Phi}_{\omega,i\tau}(X) = \int_\Sigma d^dY \left[C(X,Y,\tau)\hat{\Phi}(Y) + D(X,Y,\tau)\hat{\Pi}(Y)\right].
\end{align}
Taking expectation values of both sides with $\hat{\Phi}$ and $\hat{\Pi}$, using expression \ref{eq:freqKMS} relating expectation values to commutators, and using the canonical commutation relations, we can determine the coefficients $C,D$ in terms of two point functions,
\begin{align}\label{eq:freqbasischange}
\hat{\Phi}_{\omega,i\tau}(X) = \frac{i}{n_\omega}\int_\Sigma d^dY \left[ \langle \hat{\Phi}_{\omega,i\tau}(X) \hat{\Pi}(Y) \rangle \hat{\Phi}(Y) - \langle \hat{\Phi}_{\omega,i\tau}(X) \hat{\Phi}(Y) \rangle \hat{\Pi}(Y) \right].
\end{align}
This formula amounts to expressing the operator $\hat{\Phi}_{\omega,i\tau}(X)$ in terms of the basis of operators $\{\hat{\Phi}(X),\hat{\Pi}(X)\}$. Now we combine \ref{eq:freqbasischange} with the modular extrapolate dictionary by multiplying by $z^{-\Delta_+}$ and taking the $z\rightarrow 0$ limit. We find
\begin{align}\label{eq:modularslice}
\mathcal{O}_\omega (x,\tau) = \frac{i}{n_\omega}\int_{\Sigma} d^dY \left[ \langle \mathcal{O}_\omega(x,\tau) \hat{\Phi}(Y) \rangle \hat{\Pi}(Y) - \langle \mathcal{O}_\omega(x,\tau) \hat{\Pi}(Y)\rangle \hat{\Phi}(Y)\right].
\end{align}
This is the main result of this section, and will be key to obtaining a bulk dual for $\chi_{n,z}$.

\section{The \texorpdfstring{$\alpha$}{TEXT}-\texorpdfstring{$z$}{TEXT} divergences in holography}\label{sec:holographicdivergences}

\subsection{Bulk expression for the \texorpdfstring{$\alpha$}{TEXT}-\texorpdfstring{$z$}{TEXT} susceptibilities}\label{sec:corederivation}

We're now ready to combine some of our results to determine a bulk expression for the susceptibilities. We start with \ref{eq:tordered2ptexpression}, which we repeat here,
\begin{align}
\chi_{n,z}=\int_{-\infty}^{+\infty}d\omega F_{n,z}(\omega) \int d^d\mathbf{x} 
\,d^d\mathbf{y} \lambda(\mathbf{x})\lambda(\mathbf{y}) \langle \mathcal{O}(\mathbf{x})\mathcal{O}_{-\omega}(\mathbf{y}) \rangle, \nonumber
\end{align}
with $F_{n,z}(\omega)$ given in equation \ref{eq:Falphaz}. Now substitute for $\mathcal{O}_{-\omega}$ using expression \ref{eq:modularslice}. This yields,
\begin{align}\label{eq:unintegratedresult}
\chi_{n,z} = \int_{\Sigma}d^dX \, \int_{-\infty}^{+\infty} &d\omega \,i\frac{F_{n,z}(\omega)}{n_{-\omega}} \int d^d\mathbf{x} 
\,d^d\mathbf{y} \lambda(\mathbf{x})\lambda(\mathbf{y}) \times \nonumber \\
&\left[ \langle \mathcal{O}_{-\omega}(\mathbf{y}) \Phi(X)\rangle  \langle \mathcal{O}(\mathbf{x}) \Pi(X)\rangle - \langle \mathcal{O}_{-\omega}(\mathbf{y}) \Pi(X)\rangle \langle \mathcal{O}(\mathbf{x}) \Phi(X) \rangle \right]. 
\end{align}
To complete our calculation we need to do the integrations over $d^d\mathbf{x}$ and $d^d\mathbf{y}$. There are four of these, and we claim they are given as follows:
\begin{align}
\Phi(X)=\int d^d\mathbf{y}& \lambda(\mathbf{y}) \,\langle \mathcal{O}(\mathbf{y})\hat{\Phi}(X) \rangle, \label{eq:phiintegral} \\
\Pi(X)=\int d^d\mathbf{y}& \lambda(\mathbf{y}) \,\langle \mathcal{O}(\mathbf{y})\hat{\Pi}(X) \rangle, \label{eq:piintegral} \\
\Phi_{\omega}(X)=\int d^d\mathbf{y}& \lambda(\mathbf{y}) \,\langle \mathcal{O}_{-\omega}(\mathbf{y})\hat{\Phi}(X) \rangle, \label{eq:phiomegaintegral} \\
\Pi_{\omega}(X)=\int d^d\mathbf{y}& \lambda(\mathbf{y}) \,\langle \mathcal{O}_{-\omega}(\mathbf{y})\hat{\Pi}(X) \rangle. \label{eq:piomegaintegral}
\end{align}
We derive each of these below.

Recall expression \ref{eq:phiphihatrelationp2} which said that,
\begin{align}
\Phi(X) = \tr \left(\delta^1\rho \, \hat{\Phi}(X)\right)\nonumber
\end{align}
Using expression \ref{eq:fieldssourced} for the perturbation $\delta^1\rho$ we have
\begin{align}
\Phi(X) = \int d^d\mathbf{y} \lambda(\mathbf{y}) \,\langle \mathcal{O}(\mathbf{y})\hat{\Phi}(X) \rangle,
\end{align}
which is integral \ref{eq:phiintegral}. Taking a derivative on both sides we obtain the integral \ref{eq:piintegral}.

The two remaining integrations involve the modular frequency modes of the boundary operator $\mathcal{O}(\mathbf{x})$, so we might expect these are related to the bulk modes $\Phi_{\omega}$. These modes are defined by
\begin{align}
\Phi_{\omega}(X) = \tr \left(\delta^1\rho\, \hat{\Phi}_{\omega}(X) \right),
\end{align}
which, using \ref{eq:fieldssourced} again is just 
\begin{align}
\Phi_{\omega}(X) = \int d^d\mathbf{x} \,\lambda(\mathbf{x}) \langle \mathcal{O}(\mathbf{x}) \hat{\Phi}_{\omega}(X) \rangle.
\end{align}
Now we use the definition of the modular frequency modes along with cyclicity of the trace to rewrite this in terms of the modes of the operator $\mathcal{O}$,
\begin{align}
\Phi_{\omega}(X) &=  \int d^d\mathbf{x} \,\lambda(\mathbf{x}) \int_{-\infty}^{+\infty} ds\,e^{-is\omega} \langle \mathcal{O}(\mathbf{x}) \sigma^{-is/2\pi} \hat{\Phi}(X) \sigma^{is/2\pi} \rangle \nonumber \\
&= \int d^d\mathbf{x} \,\lambda(\mathbf{x}) \int_{-\infty}^{+\infty} ds\,e^{-is\omega} \langle \sigma^{is/2\pi} \mathcal{O}(\mathbf{x}) \sigma^{-is/2\pi} \hat{\Phi}(X)  \rangle  \nonumber \\
&= \int d^d\mathbf{x} \,\lambda(\mathbf{x}) \langle \mathcal{O}_{-\omega}(\mathbf{x}) \hat{\Phi}(X) \rangle.
\end{align}
The last equality is integral \ref{eq:phiomegaintegral}, while taking a derivative on both sides yields \ref{eq:piomegaintegral}. 

Using \ref{eq:phiintegral}-\ref{eq:piomegaintegral} to do all the boundary integrations in \ref{eq:unintegratedresult}, we arrive at
\begin{align}
\chi_{n,z} = \int_{\Sigma} d^dX \, \int_{-\infty}^{+\infty} d\omega \,i K_{n,z}'(\omega) \left[\Phi(X)\Pi_{\omega}(X)  - \Phi_{\omega}(X) \Pi(X)  \right],
\end{align}
where the function of $K_{\alpha,z}'(\omega)$ appearing is determined by \ref{eq:Falphaz} and \ref{eq:freqKMS},
\begin{align}\label{eq:modeprimed}
K_{n,z}'(\omega)=-\frac{F_{n,z}(\omega)}{n_{-\omega}} = \frac{z}{1-n} \frac{(1-e^{2\pi \omega n/z})(1-e^{2\pi \omega \frac{1-n}{z}})}{(1-e^{2\pi \omega/z})}e^{-2\pi \omega}.
\end{align}
Because the field $\Phi_{s}(X)$ satisfies the KMS condition, $\Phi_s=\Phi_{s+2\pi i}$ it's Fourier transform $\Phi_{\omega}$ is actually ambiguous up to factors of $e^{2\pi \omega}$. It's convenient then to drop the overall factor of $e^{-2\pi \omega}$ appearing in \ref{eq:modeprimed}, as this won't effect the result of the integration over modular frequency space.

Our final result then is that
\begin{align}\label{eq:mainresult}
\chi_{n,z} = \int_{\Sigma} d^dX \, \int_{-\infty}^{+\infty} d\omega \,iK_{n,z}(\omega) \left[\Phi(X)\Pi_{\omega}(X)  - \Phi_{\omega}(X) \Pi(X)\right],
\end{align}
with 
\begin{align}
K_{n,z}(\omega)= \frac{z}{1-n} \frac{(1-e^{2\pi \omega n/z})(1-e^{2\pi \omega \frac{n-1}{z}})}{(1-e^{2\pi \omega/z})}.
\end{align}
Expression \ref{eq:mainresult} gives the $\alpha$-$z$ susceptibilities purely in terms of bulk data, and constitutes the main result of this article. This expression is valid around any asymptotically AdS background and in any dimension. Notice that as a byproduct of our derivation we have that $\chi_{n,z}$ is conserved, in the sense that it is independent of the spacelike slice $\Sigma$ of the entanglement wedge chosen. This followed from the fact that we could choose any spacelike slice in expression \ref{eq:modularslice}. 

The choice $α=z=1$ corresponds to relative entropy, so our result must reduce to canonical energy in this case. To check this, it is useful to first re-write equation \ref{eq:mainresult} in an alternative form. To do this, we reinterpret the integration over modular frequency space in terms of a differential operator acting on $\Phi_s(X)$. We have
\begin{align}
\int_{-\infty}^{+\infty} d\omega\, K_{n,z}(\omega)(X) &= \int_{-\infty}^{+\infty} ds\, \Phi_s(X) \int_{-\infty}^{+\infty} d\omega\, K_{n,z}(\omega) e^{-is\omega} \\
&= \int_{-\infty}^{+\infty} ds\, \Phi_s(X) \int_{-\infty}^{+\infty} \,K_{n,z}(i\partial_s) e^{-is\omega} \\
&= \left.[ K_{n,z}(-i\partial_s)\Phi_s(X) ]\right|_{s=0}.
\end{align}
Similarly
\begin{align}
\int_{-\infty}^{+\infty} d\omega\,K_{n,z}(\omega) \Pi_\omega(X) = \left.[K_{n,z}(-i\partial_s)\Pi_s(X)]\right|_{s=0}
\end{align}
Finally, we recall that the gravitational symplectic form is defined by
\begin{align}
    W_{\Sigma}(\Phi_1,\Phi_2) \equiv \int_{\Sigma} d^dX\left[ \Phi_1(X)\Pi_2(X) - \Pi_1(X)\Phi_2(X) \right].
\end{align}
so that the susceptibilities are
\begin{align}\label{eq:mainresultoperator}
    \chi_{n,z} = W_{\Sigma}(\Phi,\hat{K}_{n,z}\Phi),
\end{align}
where $\hat{K}_{n,z}\equiv K_{n,z}(-i\partial_s)$. 

We are now able to check that our result recovers the usual relation \cite{lashkari2016canonical,faulkner2017nonlinear} between Fisher information and canonical energy when the modular flow is local and $n=1$. For $n=1$
\begin{align}
K_{n,z}(\omega) = -2\pi i\omega.
\end{align}
so that $\hat{K}_{1,1} = -2\pi \partial_s$. For near vacuum states, we can interpret the parameter $s$ as a coordinate, and define a vector field $\eta$ which points in the direction of $s$. It is conventional to choose the normalization 
\begin{align}
-2 \pi \partial_s = \mathcal{L}_\eta.
\end{align}
Then we get that
\begin{align}
    \chi_{1,1} = W_{\Sigma}(\Phi,\mathcal{L}_\eta\Phi),
\end{align}
which is the canonical energy \cite{hollands2013stability}. 

The integrated symplectic form density appearing in equation \ref{eq:mainresultoperator} is conserved in the
sense that it is invariant under deformations of the integration surface $\Sigma$. This follows because
the symplectic form is a closed differential form on spacetime when the arguments obey the equations of motion. To see that this is the case, at least for perturbations around the vacuum state, we again use that $s$ becomes a coordinate corresponding to modular flow. Further, $-2\pi \partial_s=\mathcal{L}_\eta$ is actually a Killing vector of the background metric. This implies that $\hat{K}_{n,z}(-i\partial_s)\Phi$ is a solution to the equations of motion since
\begin{align}\label{eq:solution}
    (\square_g + m^2) \hat{K}_{n,z}(-i\partial_s)\Phi = \hat{K}_{n,z}(-i\partial_s)(\square_g + m^2)\Phi = 0.
\end{align}
Since $\hat{K}_{n,z}\Phi$ is a solution to the equations of motion, we have that $\chi_{n,z} = W_{\Sigma}(\Phi,\hat{K}_{n,z}\Phi)$ independent of the choice of $\Sigma$.

\subsection{Bulk metric perturbations}\label{sec:gravityderivation}

Our derivation of bulk expressions for the boundary susceptibilities has been carried out for an insertion of a scalar operator in the boundary path integral. The boundary scalar operator sources a bulk scalar field, in terms of which we were able to express the susceptibility. We could also have considered inserting the stress tensor in the boundary path integral, which sources a bulk metric perturbation. In this case we can follow a similar derivation to arrive at a bulk expression for the susceptibility in terms of the bulk metric perturbation.

We begin again with the path integral states defined by equation \ref{eq:pathintegralstates}. Work around a background metric $G_{\mu\nu}$ and consider some metric perturbation $H_{\mu\nu}$. The bulk metric perturbation is sourced by an insertion of the stress tensor in the path integral, which leads to the state perturbation
\begin{align}
\delta^1\rho = \int d^d\mathbf{x}\,\sigma \,\lambda_{ab}(\mathbf{x})\, T^{ab}(\mathbf{x}).
\end{align}
From this we can define the modular frequency modes of the boundary perturbation,
\begin{align}
\delta^1\rho_{-\omega} = \int d^d\mathbf{x}\,\sigma\,\lambda_{ab}(\mathbf{x})\,T^{ab}_{-\omega}(\mathbf{x}).
\end{align}
Recalling expression \ref{eq:traces} for the susceptibility in terms of the state perturbation,
\begin{align}
\chi_{n,z} = \int_{-\infty}^{+\infty} d\omega F_{n,z}(\omega) \tr(\sigma^{-1}\delta^1\rho \,\delta^1\rho_{-\omega})=\int_{-\infty}^{+\infty} d\omega F_{n,z}(\omega) \tr(\sigma^{-1}\delta^1\rho_{-\omega} \,\delta^1\rho)\nonumber
\end{align}
we arrive at the spin 2 version of equation \ref{eq:tordered2ptexpression},
\begin{align}\label{eq:gravsusceptibilty}
\chi_{\alpha,z}=\int_{-\infty}^{+\infty}d\omega F_{\alpha,z}(\omega) \int d^d\mathbf{x}d^d\mathbf{y}\lambda_{ab}(\mathbf{x})\lambda_{cd}(\mathbf{y}) \langle T^{ab}(\mathbf{x})T^{cd}_{-\omega}(\mathbf{y}) \rangle.
\end{align}
Next we need the gravity statement of our change of basis formula \ref{eq:freqbasischange} and extrapolate dictionary \ref{eq:modularextrapolate2}. The extrapolate dictionary statement is
\begin{align}\label{eq:gravextrapolate}
\hat{T}_{ab}(x,\tau) = \lim_{z\rightarrow 0} z^{-\Delta_{+}} \hat{H}_{ab,i\tau}(x,z),
\end{align}
while the change of basis formula takes the general form
\begin{align}
\hat{H}_{ab,i\tau}^\omega(X) = \int_\Sigma d^dY \left( C^{ij}_{ab}(X,Y,\tau) \hat{H}_{ij}(Y) + D^{ij}_{ab}(X,Y,\tau) \hat{\Pi}_{ij}(Y) \right).
\end{align}
Note that since $\hat{H}_{ab},\hat{\Pi}_{ab}$ are symmetric in their indices we can take the two upper and two lower indices of $C$ to be symmetric, and the same for $D$.

The coefficients $C,D$ can be fixed using the relation between expectation values and commutators provided by equation \ref{eq:freqKMS}, along with appropriate canonical commutation relations,
\begin{align}
[\hat{H}_{ij}(X),\hat{\Pi}_{ab}(Y)] = \frac{i}{2}(\delta_{ia}\delta_{jb}+\delta_{ib}\delta_{ja})\delta^{(d)}(X-Y).
\end{align}
Applying also the extrapolate dictionary statement \ref{eq:gravextrapolate} we arrive at
\begin{align}
T_\omega^{ab} (y,\tau) = \frac{i}{n_\omega} \int_{\Sigma} d^dX \left[ \langle T_\omega^{ab}(y,\tau) \hat{H}_{\sigma\rho}(X) \rangle \Pi^{\sigma\rho}(X) - \langle T_\omega^{ab}(y,\tau) \hat{\Pi}_{\sigma\rho}(X)\rangle H^{\sigma\rho}(X)\right].
\end{align}
Inserting this into expression \ref{eq:gravsusceptibilty} for the susceptibility,
\begin{align}
\chi_{n,z} = \int_{\Sigma}d^dX \, &\int_{-\infty}^{+\infty} d\omega \,iK_{n,z}(\omega) \int d^d\mathbf{x}d^d\mathbf{y}\lambda_{ab}(\mathbf{x})\lambda_{cd}(\mathbf{y}) \times \nonumber \\
&\left[ \langle T_{-\omega}^{ab}(\mathbf{x}) H_{ij}(X)\rangle  \langle T^{cd}(\mathbf{y}) \Pi^{ij}(X)\rangle - \langle T_{-\omega}^{ab}(\mathbf{x}) \Pi_{ij}(X)\rangle \langle T^{cd}(\mathbf{y}) H^{ij}(X)\rangle \right].
\end{align}
Then, similar to the scalar case we use
\begin{align}
H^{ij}(X) &= \int d^d\mathbf{x}\lambda_{ab}(\mathbf{x})\langle \hat{T}^{ab}(\mathbf{x})\hat{H}^{ij}(X)\rangle, \\
\Pi^{ij}(X) &= \int d^d\mathbf{x}\lambda_{ab}(\mathbf{x})\langle \hat{T}^{ab}(\mathbf{x})\hat{\Pi}^{ij}(X)\rangle, \\
H_{\omega}^{ab}(X) &= \int d^d\mathbf{y}\lambda_{ij}(\mathbf{y}) \,\langle \hat{T}_{-\omega}^{ij}(\mathbf{y}) \hat{H}^{ab}(X) \rangle, \\
\Pi_{\omega}^{ab}(X) &= \int d^d\mathbf{y}\lambda_{ij}(\mathbf{y}) \,\langle \hat{T}_{-\omega}^{ij}(\mathbf{y}) \hat{\Pi}^{ab}(X) \rangle.
\end{align}
Using these to do the boundary integrals $\int d^d\mathbf{x}$ and $\int d^d\mathbf{y}$, we arrive at
\begin{align}
\chi_{n,z} = \int_{\Sigma} d^dX \, \int_{-\infty}^{+\infty} d\omega \,i K_{n,z}(\omega) \left[H_{ij}(X)\Pi_{\omega}^{ij}(X)  - H_{\omega}^{ij}(X) \Pi_{ij}(X)\right].
\end{align}
This reproduces our main result, expression \ref{eq:mainresult}, for state perturbations consisting of coherent excitations of the graviton field. 

\subsection{What happens away from integer \texorpdfstring{$\alpha$}{TEXT}?}\label{sec:noninteger}

To find bulk expressions for the $\alpha$-$z$ susceptibilities we made use of the two equalities in expression \ref{eq:traces}. We noted there that the second equality is true only for certain parameter values. In particular we had that 
\begin{align}
\chi_{\alpha,z} = \int_{-\infty}^{+\infty} d\omega \, F_{\alpha,z}(\omega) \tr(\sigma^{-1} \delta^1\rho\, \delta^1\rho_{-\omega}) = \int_{-\infty}^{+\infty} d\omega \, F_{\alpha,z}(\omega) \tr(\sigma^{-1} \delta^1\rho_{-\omega} \delta^1\rho) \nonumber
\end{align}
when $\alpha\in \mathbb{N}$ and $z\geq 0$, and when $\alpha \in \mathbb{R}$ and $z\in\{0,\infty\}$. 

We considered the case of $\alpha \in \mathbb{N}$ in sections \ref{sec:corederivation} and \ref{sec:gravityderivation}. However, since our two alternative expressions in \ref{eq:traces} are still true for the limiting cases $z\in\{0,\infty\}$ even for $\alpha$ non-integer, our main result \ref{eq:mainresult} is still valid in those cases. To study these cases in more detail consider the values of $K_{n,z}(\omega)$ at $z=0,\infty$,
\begin{align}
\lim_{z\rightarrow 0} K_{n,z}(\omega) &=  i \alpha \,\delta_{\omega,0}\nonumber \\
\lim_{z\rightarrow \infty} K_{n,z}(\omega) &= 2\pi i \alpha \omega.
\end{align}
By $\delta_{\omega,0}$ we mean the Kronecker delta function. From these expressions, we get that 
\begin{align}
\chi_{\alpha,0} &= 0, \nonumber \\
\chi_{\alpha,\infty} &= \alpha \,\chi_{1,1},
\end{align}
so that $\chi_{\alpha,0}$ is trivial and the bulk dual of $\chi_{\alpha,\infty}$ occurs already within the integer cases. 

What about away from the parameter region where \ref{eq:traces} is true? In general we have
\begin{align}
\chi_{\alpha,z} = \int_{-\infty}^{+\infty} d\omega \, F_{\alpha,z}(\omega) \tr(\sigma^{-1}\delta^1\rho \delta^1\rho_{-\omega}) = \int_{-\infty}^{+\infty} d\omega \, F_{\alpha,z}(\omega)e^{2\pi \omega} \tr(\sigma^{-1}\delta^1\rho_{-\omega} \delta^1\rho).
\end{align}
Repeating the steps taken at the end of section \ref{sec:holography} to write the susceptibility in terms of the time-ordered two point function, but using the above instead of \ref{eq:traces} we find
\begin{align}
\chi_{\alpha,z} = \int_{-\infty}^{+\infty} d\omega \int d^d\mathbf{x}d^d\mathbf{y}\lambda(\mathbf{x}) \lambda(\mathbf{y}) F_{\alpha,z}(\omega)e^{2\pi\omega \, \theta(x^0_E-y_E^0)} \langle \mathcal{O}(\mathbf{x})\mathcal{O}_{-\omega}(\mathbf{y}) \rangle.
\end{align}
We can now proceed as before, replacing $\mathcal{O}_\omega$ with a bulk integral according to equation \ref{eq:modularslice}. However, the appearance of the $\theta(x_E^0-y_E^0)$ dependent exponential prevents the use of equations \ref{eq:phiintegral}-\ref{eq:piomegaintegral} to do the boundary integrals. So far we have not understood how to write the susceptibilities in terms of bulk data when this factor appears. It would be interesting to do so however, in particular since the fidelity occurs at $z=\alpha=1/2$ (outside our accessible region) and there are claims in the literature that the fidelity susceptibility corresponds to a bulk volume \cite{miyaji2015distance}. 

\section{Monotonicity of the susceptibilities in the parameter \texorpdfstring{$\alpha$}{TEXT}} \label{sec:monocheck}

In this section we establish that our bulk expression for the susceptibilities satisfies Theorems \ref{thm:monotonealpharenyi} and \ref{thm:monotonealphasandrenyi} which express monotonicity in $\alpha$, at least for perturbations around the vacuum. We find that this monotonicity property holds without any assumptions placed on the bulk classical field. Consequently, we view this as a check on our result \ref{eq:mainresult}, rather than as a constraint on bulk physics.

We will give the proof for a bulk real scalar field, although the proof strategy generalizes easily. We begin by noting that the bulk field $\Phi(X,s)$ satisfies the equation of motion,
\begin{align}
(\square_{g} + m^2) \Phi(X,s) = 0.
\end{align}
We are interested in the bulk solution on the slice $\Sigma$, which we have chosen to correspond to $s=0$. Since we are working with perturbations around the vacuum, the $s$ coordinate corresponds to modular flow and
\begin{align}
\Phi(X,s) = \int_{-\infty}^{+\infty} d\omega \, e^{-i\omega s} \Phi_\omega(X).
\end{align}
We can transform our bulk equation of motion then to the modular frequency domain, for which there will be solutions labelled by the frequency $\omega$. Call these solutions $\Psi_\omega(X)$. It is convenient to give these solutions a unit normalization,
\begin{align}\label{eq:normalization}
\int_\Sigma d^dX\, \Psi_\omega^\dagger(X) \Psi_\omega(X) = 1. 
\end{align}
A given bulk solution then is a linear combination of the modes $\Psi_\omega$,
\begin{align}\label{eq:phimodes}
\Phi_\omega(X) = b(\omega)\Psi_\omega(X)\,\,\,\,\,\text{so that}\,\,\,\,\,\Phi(X,s) = \int d\omega \, e^{-is\omega} b(\omega) \Psi_\omega(X).
\end{align}
Taking a derivative with respect to $s$ on both sides yields a similar statement for the conjugate momenta,
\begin{align}\label{eq:pimodes}
\Pi_\omega(X) = -i\omega b(\omega)\Psi_\omega(X)\,\,\,\,\,\text{so that}\,\,\,\,\,\Pi(X,s) = \int d\omega \, e^{-is\omega} (-i\omega)b(\omega) \Psi_\omega(X).
\end{align}
Now consider expression \ref{eq:mainresult} for the bulk susceptibility. Using expressions 
\ref{eq:phimodes} and \ref{eq:pimodes} we arrive at
\begin{align}
\chi_{n,n} = \int_{-\infty}^{+\infty} d\omega d\omega' \,K_{n,z}(\omega)(\omega+\omega') b(-\omega) b(\omega') \int_{\Sigma} dX \,\Psi_{-\omega}(X) \Psi_{\omega'}(X).
\end{align}
The spatial integration has been isolated to the final factor written in terms of the basis of solutions to the wave equation.

Because $\Phi(X,s)$ is real and $b(\omega)$ is an arbitrary function, it follows that $b^\dagger(\omega)=b(-\omega)$ and $\Phi_\omega(X)^\dagger = \Phi_{-\omega}(X)$. Taking a complex conjugate of
\begin{align}
A_{\omega,\omega'} = \int_{\Sigma} d^dX \,\Psi_{-\omega}(X) \Psi_{\omega'}(X),
\end{align}
we see that $A_{\omega,\omega'}^* = A_{\omega',\omega}$, so that $A_{\omega,\omega'}$ can be diagonalized. Using also the normalization \ref{eq:normalization} we have imposed on $\Psi_\omega$, we find $A_{\omega,\omega'} = \delta(\omega-\omega')$. This leads to
\begin{align}
\chi_{n,z} = \int_{-\infty}^{+\infty} d\omega \,K_{n,z}(\omega) 2\omega |b(\omega)|^2.
\end{align}
Since $b(\omega)$ is arbitrary, we have that $\chi_{n,n}\leq \chi_{n+1,n+1}$ is true if and only if
\begin{align}\label{eq:inequality1}
2\omega K_{n,n}(\omega) \leq 2\omega K_{n+1,n+1}(\omega) \,\,\,\,\,\,\,\forall \,\omega \in \mathbb{R},
\end{align}
while the inequality for the Petz relative entropy amounts to\footnote{We only consider the $n=1$ case here because for $\alpha>2$ the Petz relative entropies are no longer divergences and Theorem \ref{thm:monotonealphasandrenyi} may not hold.}
\begin{align}\label{eq:inequality2}
2\omega K_{1,1}(\omega) \leq 2\omega K_{2,1}(\omega) \,\,\,\,\,\forall\,\omega \in \mathcal{R}.
\end{align}
These results establish that checking monotonicity in $\alpha$ amounts to checking an inequality between universal functions, which are in particular independent of the bulk field configuration. 

We go through the needed argument for inequality \ref{eq:inequality1}; inequality \ref{eq:inequality2} is similar. For \ref{eq:inequality1}, the explicit expression for $n\geq 2$ is
\begin{align}
2\omega\frac{n}{n-1} \frac{(e^{2\pi\omega}-1)(1-e^{-2\pi\omega \frac{n-1}{n}})}{(e^{2\pi\omega/n}-1)} \leq 2\omega\frac{n+1}{n} \frac{(e^{2\pi\omega}-1)(1-e^{-2\pi \omega \frac{n}{1+n}})}{(e^{2\pi \omega \frac{1}{n+1}}-1)}.
\end{align}
We can remove the common (everywhere positive) factor of $2\omega (e^{2\pi\omega}-1)$ to obtain the equivalent inequality
\begin{align}\label{eq:bigineq}
\frac{n}{n-1}\frac{\sinh(\frac{\pi\omega(n-1)}{n})}{\sinh(\frac{\pi \omega}{n})} \leq \frac{n+1}{n}\frac{\sinh(\frac{\pi\omega n}{n+1})}{\sinh(\frac{\pi \omega}{n+1})}.
\end{align}
To prove \ref{eq:bigineq}, we start by defining the function
\begin{align}
f(n,\omega)= \frac{n}{n-1} \sinh\left(\frac{\pi\omega(n-1)}{n}\right).
\end{align}
We can show this function is increasing with $n$ for $\omega>0$ and decreasing with $n$ for $\omega<0$. We let $u=(n-1)/n$ and take an $n$ derivative using the chain rule
\begin{align}
\frac{\partial f}{\partial n}=\frac{\partial f}{\partial u}\frac{\partial u}{\partial n} = \left[\frac{\pi\omega}{u} \cosh(\pi\omega u)-\frac{1}{u^2}\sinh(\pi\omega u) \right] \frac{1}{n^2}.
\end{align}
Define $v = \pi u \omega$ so that
\begin{align}
\frac{\partial f}{\partial n} = \left[ v\cosh(v)-\sinh (v)\right] \frac{1}{n^2 u^2}.
\end{align}
From the Taylor expansion it is straightforward to check that $\left[ v\cosh(v)-\sinh (v)\right]$ has the same sign as $v =\pi u \omega$ , so that $f(n,\omega)$ is increasing in $n$ for $\omega>0$ and decreasing in $n$ for $\omega<0$ as claimed. 

Now focus on the $\omega>0$ case. We have that $f(n-1) \leq f(n)$ so that
\begin{align}\label{eq:ineqderiv}
\frac{n}{n-1} \sinh\left(\frac{\pi\omega(n-1)}{n}\right) \leq \frac{n+1}{n} \sinh\left(\frac{\pi\omega n}{n+1}\right).
\end{align}
Also for $\omega>0$ we have
\begin{align}\label{eq:trivineq}
\sinh\left( \frac{\pi \omega}{n} \right) \geq \sinh\left( \frac{\pi \omega}{n+1} \right).
\end{align} 
Dividing inequality \ref{eq:ineqderiv} by \ref{eq:trivineq} we get the needed inequality \ref{eq:bigineq}. For $\omega<0$ the inequalities are reversed in \ref{eq:ineqderiv} and \ref{eq:trivineq}, but both sides of \ref{eq:trivineq} are negative so dividing yields \ref{eq:bigineq} again.

It is necessary to check the $n=1$ case separately by taking the $n\rightarrow 1$ limit in expression \ref{eq:inequality1}, but this is straightforward. Inequality \ref{eq:inequality2} is handled similarly.

\section{Discussion}

Our goal has been exploratory: Identify any interesting gravitational quantities still hidden in the entropy zoo. While working perturbatively around an arbitrary reference state, we have determined bulk expressions for the $\alpha$-$z$ divergences up to second order whenever $\alpha$ is an integer, as well as the limiting cases of $z\in\{0,\infty\}$ for all $\alpha \in \mathbb{R}$. We have seen explicitly that all the $\alpha$-$z$ divergences can be expressed in the bulk in terms of an operator $\hat{K}_{n,z}$ and the gravitational symplectic form,
\begin{align}
    \chi_{\alpha,z} = W_{\Sigma}(\Phi,\hat{K}_{n,z}\Phi).
\end{align}
In general $\hat{K}_{n,z}$ is a non-local operator involving an infinite number of derivatives, and it becomes local only at $n=z=1$, corresponding to the relative entropy. 

Our approach combines entanglement perturbation theory, as used for instance in \cite{faulkner2017nonlinear}, with the modular extrapolate dictionary \cite{faulkner2017bulk}. Doing so, we offer a simpler derivation that boundary Fisher information is bulk canonical energy. However, we made use of the equality of bulk and boundary modular flows which is a stronger assumption than used by previous authors. In addition to simplifying the argument showing canonical energy equals Fisher information, the stronger assumption allowed us to establish bulk duals of a larger class of objects and to work around an arbitrary background.

One consequence of our result is that, recalling our definition of the $n$-$z$ information metric \ref{eq:nzinfometric}, we find corresponding metrics on the space of bulk field perturbations. If we define,
\begin{align}
    M_{n,z}(\Phi_1,\Phi_2)\equiv \frac{1}{2}\left(W_{\Sigma}(\Phi_1+\Phi_2,\hat{K}_{n,z}(\Phi_{1}+\Phi_{2})) - W_{\Sigma}(\Phi_1,\hat{K}_{n,z}\Phi_1) - W_{\Sigma}(\Phi_2,\hat{K}_{n,z}\Phi_2)\right),
\end{align}
then from \ref{eq:mainresult} we get that $M_{n,z}(\Phi_1,\Phi_2) = \chi_{n,z}(\delta^1\rho_1,\delta^1\rho_2)$. Thus, each information metric on the space of CFT state perturbations defined by an $\alpha$-$z$ divergence is dual to a corresponding metric on the space of field perturbations in the gravitational theory. 

There are a number of directions one could take these results. First, there is an active line of research in translating constraints on information theoretic quantities into gravitational constraints \cite{banerjee2014constraining,banerjee2015nonlinear,lin2014tomography,lashkari2015inviolable,bhattacharya2015entanglement,lashkari2016canonical,lashkari2016gravitational,neuenfeld2018positive}. Using our bulk expressions for the $\alpha$-$z$ divergence allows consideration of many new constraints. For instance, recall from theorems \ref{thm:tracemonotonealphalpha} and \ref{thm:tracemonotonealphaone} that the $\alpha$-$\alpha$ and $\alpha$-$1$ divergences are monotonic under the partial trace. This immediately implies that
\begin{align}
    W_{\Sigma_A}(\Phi,\hat{K}_{n,n}^A\Phi) \leq W_{\Sigma_{AB}}(\Phi,\hat{K}_{n,n}^{AB}\Phi)
\end{align}
when $\Sigma_A\subset \Sigma_{AB}$, and we've indicated the corresponding regions in $K$ to remind the reader that this operator depends on the region. The same statement is true for $(\alpha,z)=(n,1)$. Even more simply, we obtain 
\begin{align}
    0\leq W_{\Sigma}(\Phi,\hat{K}_{n,z}\Phi)
\end{align}
because the divergences are positive. These constraints generalize those implied by monotonicity and positivity of the relative entropy \cite{lashkari2016gravitational}. It would be interesting to understand if these constraints on the $\alpha$-$z$ susceptibilities are independent of those implied by the relative entropy, and if so to express them in terms of the bulk matter stress tensor.

It would also be interesting to understand if a bulk proof of monotonicity under the partial trace can be given for the $\alpha$-$z$ susceptibilities. This understanding could be useful in the purely information theoretic context, since for certain parameter values it is not known whether the $\alpha$-$z$ divergence is monotonic \cite{carlen2018inequalities}. It is possible such a bulk understanding could be used to constrain which divergences are monotone.

A related direction would be to study black hole thermodynamics in light of our bulk expressions for the $\alpha$-$z$ divergences. Those $\alpha$-$z$ divergences which obey the data-processing inequality provide constraints on state transitions that generalize the usual second law of thermodynamics \cite{brandao2015second}. For out of equilibrium systems, this set of constraints is stronger than the single constraint provided by the second law. The possibility of using the AdS/CFT correspondence to translate these additional constraints to statements in black hole thermodynamics has been considered before \cite{bernamonti2018holographic}. Our expression \ref{eq:mainresult} seems well suited to this purpose, as it allows the calculation of many different divergences for any given bulk solution in any dimension, including black hole solutions. 

Finally, it would be interesting to try and extend our results to the case of non-integer $\alpha$ and finite, non-zero $z$. Of particular interest is the $\alpha=z=1/2$ case corresponding to the fidelity. It has been argued that the fidelity susceptibility corresponds to the volume of a bulk Cauchy slice \cite{miyaji2015distance}, and it would be interesting to give a derivation of the bulk dual of the fidelity susceptibility so as to check this expectation.

From a more general perspective, we find it interesting that all the divergences studied here translate to conserved bulk quantities written simply in terms of the symplectic form. We find this suggestive of a general mathematical connection between the information theoretic notion of divergence, conserved quantities, and symplectic structure. Additionally, the combination of entanglement perturbation theory and the modular extrapolate dictionary proved a powerful approach to studying the bulk duals of entropic quantities. It would be interesting to apply this same approach to other problems in entanglement perturbation theory. For instance, one could use this approach to study the relative entropy perturbatively to third or higher order. 

\section{Acknowledgements}

We thank Nima Lashkari, Charles Rabideau, David Wakeham, Jordan Wilson-Gerow, Dominik Neunfeld, Felix Haehl, Lampros Lamprou, Jamie Sully and Waytt Reeves for useful discussions. We also thank Mark Van Raamsdonk for guidance and many useful discussions. Eric Mintun provided an early version of the axiomatic proof given in Appendix \ref{sec:equalityofdivergences}. A discussion with Tom Faulkner produced the useful point that we should make use of statement \ref{eq:modularslice}. We also wish to thank the anonymous referee, whose input improved the clarity of our manuscript.

We acknowledge support from the It from Qubit Collaboration, which is sponsored by the Simons Foundation. AM was also supported by a CGS-D award given by the National Research Council of Canada. This research benefited from the It from Qubit summer school held at the Instituto Balserio in Bariloche, and the QINFO17 program held at The Kavli institute in Santa Barbara. Through the QINFO17 program, this research was supported in part by the National Science Foundation under Grant No. NSF PHY17-48958.

\appendix

\section{Eigenvalue expansion of the \texorpdfstring{$\alpha$}{TEXT}-\texorpdfstring{$z$}{TEXT} divergence} \label{sec:eigexpansion}

Given a reference state $\sigma$ and another state $\rho(\epsilon)$ which is perturbatively close to $\sigma$,
\begin{align}
\rho = \sigma + \epsilon \delta^1\rho + \frac{\epsilon^2}{2}\delta^2\rho + ...
\end{align}
we quoted in the main text the expression \ref{eq:knownform}, which gives the second order term of the $\alpha$-$z$ divergence in terms of the eigenvalues of the reference state and matrix elements of the perturbation. 

In this appendix we outline the derivation of that result. First, recall the identities:
\begin{align}\label{eq:iden1}
\frac{d}{dt}e^{X(t)} &= \int_0^1 dx \, e^{x X(t)}
\frac{dX(t)}{dt} e^{(1-x)X(t)},
\end{align}
\begin{align} \label{eq:iden2}
-\ln X &= \int_0^\infty \frac{ds}{s}(e^{-sX}-e^{-s}).
\end{align}
We can combine these to take derivatives of powers of density matrices,
\begin{align}
\frac{dA^p}{d\epsilon} = \frac{d}{d\epsilon} e^{p\ln A} = p\int_0^1 dx dy \int_0^\infty ds\, A^{xp} e^{-syA}\frac{dA}{d\epsilon}e^{-s(1-y)A}\,A^{(1-x)p}.
\end{align}
Second derivatives can be handled by using the product rule and continued application of our identities. 

To expand the $\alpha$-$z$ divergences define the object
\begin{align}
\tilde{\rho}_{\alpha,z} \equiv \sigma^{\frac{1-\alpha}{2z}}\rho^{\frac{\alpha}{z}}\sigma^{\frac{1-\alpha}{2z}}
\end{align}
so that
\begin{align}
D_{\alpha,z}(\rho||\sigma) = \frac{1}{\alpha-1}\log \tr((\tilde{\rho}_{\alpha,z})^z).
\end{align}
Since $(\tilde{\rho}_{\alpha,z})|_{\epsilon=0}=\sigma^{1/z}$, and $\tr(\sigma)=1$, one can check that
\begin{align}
\frac{d^2}{d\epsilon^2} D_{\alpha,z}(\rho||\sigma)_{\epsilon=0} = \frac{1}{\alpha-1} \tr\left( \frac{d^2}{d\epsilon^2} (\tilde{\rho}_{\alpha,z})^z \right).
\end{align}
We can take this second derivative of a matrix power by repeated application of the identities \ref{eq:iden1} and \ref{eq:iden2}. At each use of either of the matrix identities, we introduce an additional integration parameter. Our final expression involves 12 integrals, 8 from uses of \ref{eq:iden1} and $4$ from uses of \ref{eq:iden2}. We find an expression which is bilinear in the first order perturbation,
\begin{align}\label{eq:unintegratedtrace}
\chi_{\alpha,z} = \int_0^1 dx_1 ... dx_8 \int_0^\infty ds_1...ds_4 \,\tr(A_{\alpha,z}(\sigma,x_i,s_i) \delta^1 \rho B_{\alpha,z}(\sigma,x_i,s_i)\delta^1\rho).
\end{align}
The form of the functions of operators $A,B$ can be worked out explicitly by carrying through the application of identities \ref{eq:iden1} and \ref{eq:iden2}. 

It's possible to do the $x_i$ and $s_i$ integrals. We consider the identity operator
\begin{align}
\mathcal{I} = \int d\sigma_a \,\ketbra{\sigma_a}{\sigma_a},
\end{align}
where the states $\ket{\sigma_a}$ form an eigenbasis of $\sigma$, so that $\sigma \ket{\sigma_a} = \sigma_a \ket{\sigma}$. Insert this identity operator in \ref{eq:unintegratedtrace} twice to find
\begin{align}
\chi_{\alpha,z} = \int d\sigma_a d\sigma_b \left(\int_0^1 dx_1...dx_8 \int_0^\infty ds_1...ds_4 A_{\alpha,z}(\sigma_a,x_i,s_i)B_{\alpha,z}(\sigma_b,x_i,s_i) \right)|\delta^1\rho_{ab}|^2.
\end{align}
Now the integrals over the operator valued functions $A,B$ are just integrals over ordinary functions and can be done explicitly. The result is
\begin{align}
\chi_{\alpha,z}=\frac{d^2}{d\epsilon^2} D_{\alpha,z}(\rho||\sigma)|_{\epsilon=0} = \frac{z}{1-\alpha}\int d\sigma_a d\sigma_b\frac{(\sigma_a^{\alpha/z}-\sigma_b^{\alpha/z})(\sigma_a^{\frac{1-\alpha}{z}}-\sigma_b^{\frac{1-\alpha}{z}})}{(\sigma_a-\sigma_b)(\sigma_a^{1/z}-\sigma_b^{1/z})} |\delta^1\rho_{ab}|^2
\end{align}
as quoted in the main text. 

As a check, we can take the limit as $\alpha=z\rightarrow 1$ to find
\begin{align}
\chi_{1,1} = \int d\sigma_a d\sigma_b\frac{\ln\sigma_a-\ln\sigma_b}{\sigma_a-\sigma_b} |\delta^1\rho_{ab}|^2.
\end{align}
This expression agrees with \cite{faulkner2017nonlinear} (see their equation B.6\footnote{They used a discrete sum over eigenvalues of $\sigma$ where more properly they should have an integral, as we used here. The distinction becomes important in treating the modular frequency transformation, in particular the Dirac delta function appearing in \ref{eq:modfrequencyfirstform} only makes sense inside an integral over eigenvalues.}), where they used this to relate Fisher information and canonical energy.

\section{Determining where \texorpdfstring{$\tilde{F}_{\alpha,z}(s)$}{TEXT} is even} \label{sec:Fevenproof}

Recall that we defined
\begin{align}
\tilde{F}_{\alpha,z}(s)=\int_{-\infty}^{+\infty} d\omega F_{\alpha,z}(\omega)e^{is\omega},
\end{align}
where
\begin{align}\label{eq:fomega}
F_{n,z}(\omega) = \frac{z}{1-\alpha} \frac{(1-e^{2\pi \omega\alpha/z})(1-e^{2\pi \omega \frac{1-\alpha}{z}})}{(1-e^{2\pi \omega})(1-e^{2\pi \omega/z})}.
\end{align}
We would like to understand where in the $\alpha$-$z$ parameter space $\tilde{F}_{\alpha,z}(s)$ is even. We prove here that $\tilde{F}_{\alpha,z}(s)$ is even when $\alpha$ is an integer and $z \geq 0$, and for $\alpha\in \mathbb{R}$ when $z\in\{0,\infty\}$. 

From \ref{eq:fomega} it follows that $F_{\alpha,z}(-\omega) = e^{2\pi \omega} F_{\alpha,z}(\omega)$. This gives
\begin{align}\label{eq:ft}
\tilde{F}_{\alpha,z}(-s) = \int_{-\infty}^{+\infty} d\omega\, e^{2\pi \omega} F_{\alpha,z}(\omega) e^{ i s\omega}.
\end{align}
Its possible to do the $d\omega$ integral by closing the contour in a semicircle at infinity, either in the upper or lower half plane depending on the sign of $s$. From \ref{eq:fomega}, we see that $F_{\alpha,z}(\omega)$ has poles at 
\begin{align}
\omega = in \,\,\,\,\, \text{and}\,\,\,\,\, \omega = i n z.
\end{align}
There are a few cases to consider. 

\subsection*{Case 1: $0<z<\infty$ and $z$ is irrational}

There are only first order poles. The integral in equation \ref{eq:ft} is a sum over residues, 
\begin{align}
2\pi i\sum_{n} \frac{z}{1-\alpha} (1-e^{2\pi in\alpha/z})(1-e^{2\pi in \frac{1-\alpha}{z}}) e^{2\pi in} + 2\pi i\sum_{n} \frac{z}{1-\alpha} (1-e^{2\pi in \alpha })(1-e^{2\pi in (1-\alpha)}) e^{2\pi i n z}.
\end{align}
We can see that if $\alpha \in \mathbb{N}$ the residues from the poles at $inz$ are zero. Only the poles at $\omega=in$ contribute. In that case, the factor of $e^{2\pi \omega} = e^{2\pi i n}$ is always 1, so we can leave it out of the integral, 
\begin{align}
\tilde{F}_{n,z}(-s) = \int_{-\infty}^{+\infty} d\omega\,  e^{2\pi \omega} F_{\alpha,z}(\omega) e^{-is\omega} = \int_{-\infty}^{+\infty} d\omega\,  F_{\alpha,z}(\omega) e^{-i s\omega} = \tilde{F}_{n,z}(s)
\end{align}
so that $\tilde{F}_{n,z}(s)$ is even.

\subsection*{Case 2: $0<z<\infty$ and is rational}

There will be double poles coming from $\omega = in = imz$ for $z = n/m$, as well as additional simple poles. For the simple poles the same argument as in case 1 shows that when $\alpha$ is an integer we may ignore the factor of $e^{2\pi \omega}$ in \ref{eq:ft}. The double poles contribute the following sum of residues to the contour integral,
\begin{align}\label{eq:doublepoles}
\frac{z}{1-\alpha} \sum_{n,m} \left((\partial_\omega e^{2\pi \omega}) (1-e^{2\pi \omega \alpha/z})(1-e^{2\pi \omega \frac{1-\alpha}{z}}) + e^{2\pi \omega} \partial_{\omega}((1-e^{2\pi \omega \alpha/z})(1-e^{2\pi \omega \frac{1-\alpha}{z}})) \right)|_{\omega = imz = in}.
\end{align}
If the $e^{2\pi \omega}$ had not been present, the contribution would have been
\begin{align}
\frac{z}{1-\alpha} \sum_{n,m} \left( e^{2\pi \omega} \partial_{\omega}((1-e^{2\pi \omega \alpha/z})(1-e^{2\pi \omega \frac{1-\alpha}{z}})) \right)|_{\omega = imz = in}.\nonumber 
\end{align}
We can see that these coincide when $z = imz = in$, since then the first term in equation \ref{eq:doublepoles} is zero and in the second we have again $e^{2\pi i n}= 1$.

\subsection*{Case 3: $z=0$}

It can be checked that 
\begin{align}
\lim_{\omega\rightarrow 0} F_{\alpha,z}(\omega) &= \alpha \nonumber \\
\lim_{z\rightarrow 0} F_{\alpha,z}(\omega\neq 0) &= 0,
\end{align}
which establishes that $F_{\alpha,z}(\omega)$ becomes a Kronecker delta function for $z\rightarrow 0$. Then $\tilde{F}_{\alpha,z}(s)$ vanishes and so is even. Note that in this $z\rightarrow 0$ case we do not need $\alpha$ to be an integer, but since $\tilde{F}_{\alpha,z}(s)=0$ we have that $\chi_{\alpha,0}=0$, so this case is trivial.

\subsection*{Case 4: $z\rightarrow \infty$}

Taking the limit,
\begin{align}
\lim_{z\rightarrow \infty} F_{\alpha,z}(\omega) = \frac{2\pi \alpha \,\omega}{e^{2\pi \omega}-1}
\end{align}
we see that $F_{\alpha,\infty}(\omega)$ has poles only at $\omega=in$, $n \in \mathbb{N}$, so that $\tilde{F}_{\alpha,\infty}(s)$ is even. 

Finally, note that for $D_{\alpha,z}(\rho||\sigma)$ to be a divergence we need $z \geq |\alpha-1| \geq 0$, so these four cases are comprehensive.

\section{Equality of bulk and boundary divergences}\label{sec:equalityofdivergences}

The $\alpha$-$\alpha$ divergences and $\alpha$-$1$ divergences have operational meanings in terms of quantum hypothesis testing \cite{mosonyi2015quantum}. Roughly speaking, they measure how difficult it is to distinguish the density matrix $\rho$ from the density matrix $\sigma$. Given this operational meaning and the notion of subregion-subregion duality in AdS/CFT we might expect that the $\alpha$-$z$ divergences in the bulk and boundary are equal,
\begin{align}\label{eq:newJLMS}
D_{\alpha,z}(\rho_R||\sigma_R) = D_{\alpha,z}(\rho_{W}||\sigma_{W}),
\end{align}
where $R$ is a boundary subregion and $W$ is the corresponding bulk entanglement wedge. The above was already argued for by JLMS \cite{jafferis2016relative} in the special case where $\alpha=z=1$. We can verify that the more general statement \ref{eq:newJLMS} is indeed the case, at least under certain assumptions on the relation between $\rho$ and $\sigma$.

First suppose that we have two states $\rho$, $\sigma$ which we assume share an area operator,
\begin{align}\label{eq:areaops}
H_{\rho_R} &= \frac{\hat{A}}{4G} + H_{\rho_{W}} \nonumber \\
H_{\sigma_R} &= \frac{\hat{A}}{4G} + H_{\sigma_{W}}.
\end{align}
Recall also that these expressions should be understood inside of expectation values. Consider the definition of the $\alpha$-$z$ divergence,
\begin{align}
D_{\alpha,z}(\rho_R||\sigma_R) = \frac{1}{\alpha-1} \log \tr \left(\rho_R \rho_R^{-1}[\sigma_R^{\frac{1-\alpha}{2z}} \rho_R^{\frac{\alpha}{z}} \sigma_R^{\frac{1-\alpha}{2z}}]^z\right) = \frac{1}{\alpha-1} \log \langle \rho_R^{-1}[\sigma_R^{\frac{1-\alpha}{2z}} \rho_R^{\frac{\alpha}{z}} \sigma_R^{\frac{1-\alpha}{2z}}]^z\rangle \nonumber 
\end{align}
where we have inserted the identity $\mathcal{I}=\rho_R\rho_R^{-1}$. Writing $\rho = e^{-H_{\rho_R}}$, $\sigma = e^{-H_{\sigma_R}}$ and using \ref{eq:areaops} we can relate this to the bulk divergence. Because the area operator commutes with the modular Hamiltonians of both $\rho$ and $\sigma$, we can collect all the factors with $\hat{A}$ and find that they cancel. We are left with the result that
\begin{align}
D_{\alpha,z}(\rho_R||\sigma_R) = D_{\alpha,z}(\rho_W||\sigma_W),
\end{align}
which reduces to the JLMS result for $\alpha=z=1$. 

There is another way to arrive at the same result which we find appealing. In the understanding of entanglement wedge reconstruction in terms of quantum error correction, Harlow \cite{harlow2017ryu} argued that the bulk and boundary density matrices are related according to
\begin{align}
\rho_R = U \left( \bigoplus_\alpha\left(\rho^\alpha \otimes \chi^\alpha \right) \right) U^\dagger,
\end{align}
where $R$ labels the subregion being considered, $\rho_R$ is the boundary state, and $U$ is a unitary. The density matrices $\rho^\alpha$ are not individually normalized but satisfy $\sum_\alpha \tr(\rho^\alpha)=1$. The state
\begin{align}\label{eq:defrhochi}
\rho_W \equiv \bigoplus \rho^\alpha 
\end{align}
is the bulk density matrix, while the $\chi^\alpha$ fix the area term in the Ryu-Takayanagi formula.

In this language, two boundary states having the same area operator means they share the $\chi^\alpha$'s,
\begin{align}\label{eq:definitionstates}
\rho_R = U \left( \bigoplus_\alpha\left( \rho^\alpha \otimes \chi^\alpha \right) \right) U^\dagger, \nonumber \\
\sigma_R = U \left( \bigoplus_\alpha\left( \sigma^\alpha \otimes \chi^\alpha \right) \right) U^\dagger.
\end{align}
Then, we can apply the axioms that define a divergence to show that $D(\rho_R||\sigma_R)=D(\rho_W||\sigma_W)$. We give the steps below, indicating which of the axioms from section \ref{sec:divergences} or definitions from this section we use at each stage.
\begin{align}
D(\rho_R||\sigma_R) &= D\left( \bigoplus_\alpha\left( \rho^\alpha \otimes \chi^\alpha \right) ||  \bigoplus_\alpha\left( \sigma^\alpha \otimes \chi^\alpha \right) \right) \,\,\,\,\,\,\,\,\,\text{(definition \ref{eq:definitionstates}, unitary invariance)} \nonumber \\
&= \sum_\alpha \tr(\rho_\alpha) g\left( D(\rho_\alpha\otimes \chi_\alpha||\sigma_\alpha\otimes \chi_\alpha) \right) \,\,\,\,\,\,\,\,\,\,\,\,\,\text{(generalized mean value)} \nonumber \\
&= \sum_\alpha \tr(\rho_\alpha) g\left( D(\rho_\alpha ||\sigma_\alpha)+D(\chi_\alpha||\chi_\alpha) \right) \,\,\,\,\,\,\,\text{(additivity)} \nonumber \\
&=\sum_\alpha \tr(\rho_\alpha) g\left( D(\rho_\alpha ||\sigma_\alpha) \right) \,\,\,\,\,\,\,\,\,\,\,\,\,\,\,\,\,\,\,\,\,\,\,\,\,\,\,\,\,\,\,\,\,\,\,\,\,\,\,\,\,\,\text{(order)}\nonumber \\
&= D(\bigoplus \rho_\alpha||\bigoplus \sigma_\alpha) \,\,\,\,\,\,\,\,\,\,\,\,\,\,\,\,\,\,\,\,\,\,\,\,\,\,\,\,\,\,\,\,\,\,\,\,\,\,\,\,\,\,\,\,\,\,\,\,\,\,\,\,\,\,\,\, \text{(generalized mean value)} \nonumber \\
&= D(\rho_W||\sigma_W) \,\,\,\,\,\,\,\,\,\,\,\,\,\,\,\,\,\,\,\,\,\,\,\,\,\,\,\,\,\,\,\,\,\,\,\,\,\,\,\,\,\,\,\,\,\,\,\,\,\,\,\,\,\,\,\,\,\,\,\,\,\,\,\,\,\,\,\,\,\,\,\,\, \text{(definition \ref{eq:defrhochi})}
\end{align}
Since there may be quantum divergences which are not $\alpha$-$z$ divergences, this proof is more general than the one using the relation between bulk and boundary modular Hamiltonians. 

\bibliographystyle{unsrt}
\bibliography{biblio}

\end{document}